\documentclass[aps,prb,reprint,twocolumn,superscriptaddress,showpacs,floatfix]{revtex4-1}

\usepackage{hyperref}
\usepackage{graphicx}
\usepackage{amsmath}
\usepackage{amssymb}
\usepackage{bbm}
\usepackage{dashrule}
\usepackage{subfigure}
\usepackage{enumitem}
\usepackage{color}
\begin{document}

\title{Coupled Cluster as an impurity solver for Green's function embedding methods}
\author{Avijit Shee}
\affiliation{Department of Chemistry, University of Michigan, Ann Arbor, Michigan 48109, USA}
\author{Dominika Zgid}
\affiliation{Department of Chemistry, University of Michigan, Ann Arbor, Michigan 48109, USA}
\date{\today}
\begin{abstract}
We investigate the performance of Green's function coupled cluster singles and doubles (CCSD) method as a solver for Green's function embedding methods. 
To develop an efficient CC solver, we construct the one-particle Green's function from the coupled cluster (CC) wave function based on a non-hermitian Lanczos algorithm. The major advantage of this method is that its scaling does not depend on the number of frequency points. We have tested the applicability of the CC Green's function solver in the weakly to strongly correlated regimes by employing it for a half-filled 1D Hubbard model projected onto a single site impurity problem and a half-filled 2D Hubbard model projected onto a 4-site impurity problem. For the 1D Hubbard model, for all interaction strengths, we observe an excellent agreement with the full configuration interaction (FCI) technique, both for the self-energy and spectral function. For the 2D Hubbard, we have employed an open-shell version of the current implementation and observed some discrepancies from FCI in the strongly correlated regime.
Finally, on an example of a small ammonia cluster, we analyze the performance of the Green's function CCSD solver within the self-energy embedding theory (SEET) with Hartee-Fock (HF) and Green's function second order (GF2) for the treatment of the environment.
\end{abstract}
\maketitle

\section{Introduction}
Coupled cluster (CC) methods~\cite{cizek1966, bartlett:reviewcc} are extremely successful in theoretical chemistry since they can reach spectroscopic accuracy for weakly~\cite{piecuch2018,stoneburner2017} and moderately correlated problems. The popularity of CC methods stems from unique properties of the CC exponential ansatz that (i) allows for a generation of some of the higher level electron excitations in terms of lower level excitation (ii) is both size consistent and size extensive for standard versions of the CC theory (iii) yields accurate  methods with only polynomial scaling, eg. coupled clusters singles and doubles with perturbative triples CCSD(T) that scales only as $O(n^7)$, where $n$ is the number of orbitals.

Motivated by the efficiency of the CC ansatz for finite molecular problems at zero temperature, we investigate its performance as an impurity solver for the Green's function embedding methods. We believe that such a solver would be appealing to both the quantum chemistry as well as condensed matter community. 

For quantum chemists, the CC impurity solver applied to Green's function embedding methods would allow for an easy extension of multiple molecular CC versions to periodic problems without any burden of an actual implementation in the momentum space. While multiple very accurate CC methods with various  approximations of the excitation level contained in the cluster operator were developed at a molecular level~\cite{Kallay2000,Matthews2015}, CC applications to periodic systems~\cite{Gruneis2011, Booth2012, Gruneis2018, HIRATA2001, Hirata2004} remained limited to a simple crystalline problems mostly due to a large cost of a brute force calculation on the entire problem. Moreover, these CC periodic implementations usually yielded the electronic energy of a unit cell and the periodic equation of motion coupled cluster method (EOM-CC)~\cite{HIRATA2001, Hirata2004, katagiri2005, Mcclain2017} was implemented in even fewer cases. The implementation of the Green's function CC solver would allow to easily obtain spectral information for solids. This is especially important since a theoretically evaluated spectral function provides very valuable information about the electronic structure of the problem since it can be directly compared to experimental data.

For condensed matter community, the CC solver could provide an alternative to explicit Hamiltonian representation solvers used for the dynamical mean field theory (DMFT)~\cite{Georges96,Georges92,Georges04}. 
For these solvers, the bath orbitals are treated explicitly since the hybridization function describing the coupling between the impurity and environment is discretized into a finite set of bath orbitals. Consequently, the resulting Anderson impurity model (AIM) is a finite problem.
Most current solvers working in an explicit Hamiltonian representation are based on the exact diagonalization (ED) scheme ~\cite{Caffarel94,PhysRevB.96.235149}, its truncated versions~\cite{ZgidPRB2012,Zgid11}, or density matrix renormalization group (DMRG)~\cite{DMRG_solver2014}. Both ED and its truncated versions are limited to a small number of impurity and bath orbitals due to their exponential scaling. CC methods scaling is only polynomial and they are able to treat problems that exceed 200 orbitals, thus, giving access to much larger impurity problems than the ones commonly employed today.

In this paper, we implement an evaluation of the CC Green's function using both the restricted and unrestricted coupled cluster singles and doubles (RCCSD and UCCSD) methods based on the Lanczos procedure. The details of this procedure, which allows us to avoid an explicit frequency dependence and can be equally easily executed on the real and imaginary axis, are described in Sec.~\ref{sec:method}- \ref{sec:implementation}. 
Let us mention here that many implementations of CCGF exist in the literature. The earliest are from Nooijen and coworkers~\cite{NooijenIJQC1993,NooijenIJQC1992} while the most recent ones can be found in Ref.~\onlinecite{KowalskiPRA2016,KowalskiJCP2014,KowalskiJCP2014,JCTCPeng2018,Mcclain2017}. These solve for the Green's function using linear equations on the real axis. While the evaluation of GFCC was known in molecular chemistry, its use as a solver for embedding techniques was not investigated before. We are aware of only one paper by Tianyu Zhu et al.~\cite{GFCC_Garnet} that is concurrent to our work that investigates this issue.

Subsequently, in Sec.~\ref{sec:2d_hubbard}-\ref{sec:1D Hubbard}, we test the performance of both RCCSD and UCCSD solvers on impurity problems that 1D and 2D Hubbard models can be mapped onto.  The CCSD self-energies are then compared with the ones coming from the full configuration interaction (FCI) as well as some of its truncated versions. In Sec.~\ref{sec:seet_gfcc}, we describe in detail how to employ the CC Green's function as a solver for the self-energy embedding theory (SEET)~\cite{Zgid15,Tran15b,Tran16,zgid_njp17,Tran_generalized_seet,GWSEET2017,simons_benchmark2,Tran_useet,seet_periodic1} and we test it on an ammonia cluster employing both HF and GF2~\cite{Dahlen05,Zgid14,Rusakov16,Phillips15,Kananenka15,Kananenka16,Welden16,kananenka_hybrif_gf2,Iskakov_Chebychev_2018} for the treatment of the environment.  In the SEET framework, we analyze electronic energies, self-energies, and spectra. 
Finally, we form our conclusions in Sec.~\ref{sec:conclusion}.

\section{Method}\label{sec:method}

The one-particle Green's function is expressed as 
\begin{align}
G_{pq}(\omega) = {} & G^e_{pq}(\omega)+G^h_{pq}(\omega) \nonumber \\
G^h_{pq}(\omega) = {} & \langle \Psi | a_p^\dagger \frac{1}{\omega + \mu + (H - E_{gr}) - i\eta } a_q | \Psi \rangle  \nonumber \\
G^e_{pq}(\omega) = {} &  \langle \Psi | a_p \frac{1}{\omega + \mu - (H - E_{gr}) + i\eta}  a_q^\dagger | \Psi \rangle \label{Eq:gf},
\end{align}
where $|\Psi\rangle$ is the ground state wave function and  $p$ and $q$ are labeling orbitals. The fermionic annihilation and creation operators are defined as $a_p$  and $a_p^\dagger$. The superscripts $e$ and $h$ stand for electron (electron affinity) and hole (ionization potential), respectively.

In general, the Hamiltonian $H$ in Eq.~\ref{Eq:gf} is  defined as 
\begin{align}
H=\sum_{pq}t_{pq} a_p^\dagger a_q + \frac{1}{4}\sum_{pqrs} v_{pqrs} a_p^\dagger a_q^\dagger a_s a_r.
\label{Eq:ham}
\end{align}
The one-body Green's function can be evaluated either on the real or imaginary axis.  One the real axis, the frequency $\omega$ is a real number and a small value of broadening, $\eta$, is added to assure that resulting peaks have a finite width. On the imaginary axis, the frequency $\omega$ is an imaginary number and a commonly used discretization is in terms of inverse temperature $\beta=1/(k_BT)$, where $k_B$ is the Boltzmann constant. Such a discretization results in a Matsubara frequency grid with imaginary frequency points defined as $\omega_n=i(2n+1)\pi/\beta$.

In the CC method, the ground state wave function is obtained by applying a wave operator $e^T$ to an approximate ground state wave function, for example, the Hartree-Fock determinant $|\Phi_0\rangle$. The cluster operator, $T$, induces various ranks of hole-particle excitations and the ground state CC wave function is expressed as $|\Psi\rangle = e^T |\Phi_0\rangle$. By increasing the rank of the $T$ operator one can systematically reach the exact solution limit, however, that incurs huge computational expense. Therefore, in many practical applications $T$ is usually truncated at the single and double excitation level
\begin{equation}
 T = \sum_{i,a} t_i^a a_a^\dagger a_i + \frac{1}{4} \sum_{i,j,a,b} t^{ab}_{ij} a_a^\dagger a_b^\dagger a_j a_i \label{Eq:amplitude}. 
\end{equation}
The single and double excitations amplitudes, $t_i^a$ and $t^{ab}_{ij}$, respectively  are evaluated by first similarity transforming $H$ to $\widetilde{H} = e^{-T} H e^{T}$, and then projecting up-to doubly excited determinants ($\langle \chi |$) in the Schr\"{o}dinger equation
\begin{align}
\langle \chi_i^a | \widetilde{H} | \Phi_0 \rangle = {}&  0 \\
\langle \chi_{ij}^{ab} | \widetilde{H} | \Phi_0 \rangle = {}&  0.
\end{align}
The ket-side of the transformed Hamiltonian is well-defined by $e^T |\Phi_0 \rangle$. However, the bra-wavefunction is not an adjoint of the ket part, since $e^T$ is not an unitary operator due to $T^\dagger \neq -T$. Therefore the transformed Hamiltonian is not hermitian. Consequently, one defines a bi-orthogonal bra obeying the following criterion
\begin{equation}
\langle \Psi_L | \Psi_R \rangle = 1.
\end{equation}

In the equation above one distinguishes between the right and left hand CC wave function by designating them $|\Psi_R \rangle$ and $\langle \Psi_L|$, respectively. $\langle \Psi_L |$ can be defined in many different ways \cite{}, however, here we take the most common choice and we express  $\langle \Psi_L |$ in terms of linear de-excitation operator $\Lambda$ \cite{Koch1990} as
\begin{equation}
\langle \Psi_L | = \langle \Phi_0 | (1+\Lambda) e^{-T}.
\end{equation}
Similar to the $T$ operator, $\Lambda$ is  truncated as well at the singles-doubles de-excitation level
\begin{equation}
 \Lambda = \sum_{i,a} l_a^i a_i^\dagger a_a + \frac{1}{4} \sum_{i,j,a,b} l^{ij}_{ab} a_i^\dagger a_j^\dagger a_b a_a \label{lambda}. 
\end{equation}

Now after inserting the definition of $|\Psi_R \rangle$ and $\langle \Psi_L|$ into Eq. \ref{Eq:gf} one obtains \cite{NooijenIJQC1993, NooijenIJQC1992}
\begin{widetext}
\begin{equation}
G^{CC}_{pq}(\omega) =  \langle \Phi_0 | (1+ \Lambda) e^{-T} a_p^\dagger \frac{1}{\omega + \mu + (H - E_{gr}) - i\eta } a_q e^T | \Phi_0 \rangle + 
                                                      \langle \Phi_0 | (1+ \Lambda) e^{-T} a_p \frac{1}{\omega + \mu - (H - E_{gr}) + i\eta}  a_q^\dagger e^T | \Phi_0 \rangle \label{Eq:ccgf1}.
\end{equation}
\end{widetext}
By applying unity, $1 = e^T e^{-T}$ in Eq. \ref{Eq:ccgf1}, the expression can be further reduced to 

\begin{align}
G^{CC}_{pq} (\omega) = {} & \langle \Phi_0 | (1+ \Lambda) \overline{a_p^\dagger} \frac{1}{\omega + \mu + \overline{H} - i\eta } \overline{a}_q| \Phi_0 \rangle + \nonumber \\
                                                    {} &  \langle \Phi_0 | (1+ \Lambda) \overline{a}_p \frac{1}{\omega + \mu - \overline{H} + i\eta}  \overline{a_q^\dagger} | \Phi_0 \rangle \label{Eq:ccgf2},
\end{align}
where, $\overline{a}_p = e^{-T} a_p e^T$, $\overline{a_p^\dagger} = e^{-T} a_p^\dagger e^T$ and $\overline{H} = \widetilde{H} - E_{gr}$. 

\section{Solving CCGF via Linear Equation Solver} \label{sec:Lsolver}

To evaluate CC Green's function (CCGF) from Eq. \ref{Eq:ccgf2}, we first define two equations \cite{KowalskiJCP2014, KowalskiPRA2016}
\begin{align}
(\omega + \mu + \overline{H} - i\eta ) X_p | \Phi_0 \rangle ={} &  \overline{a}_p |\Phi_0 \rangle, \label{Eq:leq1} \\
(\omega + \mu - \overline{H} + i\eta) Y_p | \Phi_0 \rangle ={} & \overline{a_p^\dagger} |\Phi_0 \rangle \label{Eq: leq2}.
\end{align}
containing auxiliary quantities $X_p$ and $Y_p$. Using these quantities 
the CCGF expression can be rewritten as
\begin{align}
G^{CC}_{pq} (\omega) ={} & \langle \Phi_0 |  (1+ \Lambda) \overline{a_p^\dagger} X_q | \Phi_0 \rangle + \nonumber \\
                      {} & \langle \Phi_0 |  (1+ \Lambda) \overline{a}_p Y_q | \Phi_0 \rangle, \label{Eq:ccgf3}
\end{align}
where $X_q$ and $Y_q$ in Eq.\ref{Eq:ccgf3} are evaluated by solving simultaneous linear equations defined in Eqs.~\ref{Eq:leq1} and \ref{Eq: leq2}. These equations are frequency dependent and for imaginary frequencies are complex. 
In Eqs.~\ref{Eq:leq1} and \ref{Eq: leq2}, $\overline{a}_p$ and $\overline{a_p^\dagger}$ lead to a naturally truncated Baker-Campbell-Housdorff (BCH) expansion with only linear power terms remaining 
\begin{align}
\overline{a}_p ={} & a_p + [a_p, T], \label{Eq:transap}\\
\overline{a_p^\dagger} ={} & a_p^\dagger + [a_p^\dagger, T].
\end{align}
Consequently, with the CCSD truncation of amplitudes $T$, the operator $\overline{a}_p$ results only in 1h or 2h-1p excitation type and  operator $\overline{a}_p^\dagger$ yields only  1p and 2p-1h excitations. In practice, we choose vectors which correspond to 1h, 2h-1p excitations for IP and 1p, 2p-1h type excitations for EA type Green's function to solve Eqs.~\ref{Eq:leq1} and \ref{Eq: leq2}.  

\section{Lanczos Procedure} \label{sec:lanczos}

The procedure described in Sec. \ref{sec:Lsolver} avoids the explicit calculation of matrix inversion by introducing frequency dependent quantities such as $X_p$ and $Y_p$ and solving a linear set of equations. In this procedure, one needs to solve a separate set of equations for every frequency point. We can, however, calculate the inversion, which does not scale proportionately with the number of frequency points by means of the Lanczos procedure \cite{golub-1996}. 

In the Lanczos procedure, a special basis is constructed in which the Hamiltonian is tridiagonal. In our case, since the Hamiltonian is non-hermitian, we have different sets of left- and right-hand side of Lanczos chain vectors. We call them $P$ and $Q$, respectively. We construct them in such a way that they obey the bi-orthogonality relationship, $P^TQ = 1$. In the $P$ and $Q$ basis, a tridiagonal matrix $T$ is expressed as
\begin{equation}
T = P^T \overline{H} Q =   \begin{pmatrix} 
      \alpha_1 & \gamma_1 & 0 & ... & 0 \\
      \beta_1 & \alpha_2 & \gamma_2 & ... & 0 \\
      0       &  \beta_2 & \alpha_3 & \ddots & 0  \\
      0        &  0       & \ddots  & \ddots & 0 \\
   \end{pmatrix} \label{Eq:tridiagonal}
.
\end{equation}
Using the bi-orthogonality condition, we can derive the recursion relations from Eq. \ref{Eq:tridiagonal} for $P$ and $Q$ as $\overline{H}Q = QT$ and $TP^T = P^T\overline{H}$. In terms of columns of the $P$ and $Q$ matrices, denoted here as $\{p_j\}$ and $\{q_j\}$, those recursion relations at i-th iteration are
\begin{eqnarray}
\overline{H} q_i = \gamma_{i-1}q_{i-1} + \alpha_i q_i + \beta_i q_{i+1}, \\
p_i^T \overline{H} = \beta_{i-1} p_{i-1}^T + \alpha_i p_i^T + \gamma_ip_{i+1}^T. 
\end{eqnarray}

If we assume  $\gamma_0q_1 = 0$ and $\beta_0p_1^T = 0$, at (i+1)th iteration, then we get
\begin{eqnarray}
q_{i+1} = \frac{\overline{H}q_i - \gamma_{i-1}q_{i-1} - \alpha_i q_i} {\beta_i} = \frac{r_i}{\beta_i}, \label{Eq:rightLC} \\
p_{i+1}^T = \frac{p_i^T\overline{H} - \beta_{i-1} p_{i-1}^T - \alpha_ip_i^{T}}{\gamma_i} = \frac{s_i^T}{\gamma_i}. \label{Eq:leftLC}
\end{eqnarray}
From Eqs. \ref{Eq:rightLC} and \ref{Eq:leftLC}, we can uniquely define $\alpha_i = p_i^T \overline{H} q_i$. On the other hand, there is no unique definition of $\beta_i$ and $\gamma_i$ that we can choose. However, they  follow the relationship $\gamma_i \beta_i = s_i^T r_i$ (using, $P^TQ = 1$), which allows us to choose $\beta_i = |s_i^Tr_i|^{1/2}$, and hence $\gamma_i = (s_i^Tr_i)\beta_i^{-1}$. A possible alternative definition is $\beta_i = |r_i^Tr_i|^{1/2}$, hence $\gamma_i = (s_i^Tr_i)\beta_i^{-1}$. With these choices of $\beta_i$ and $\gamma_i$, we can fix the bi-orthogonality between the Lanczos chains at every iteration. We notice here that $\beta_i$ and $\gamma_i$ can only differ by a sign factor within the first definition. The convergence of a Lanczos procedure is typically achieved when the off-diagonal elements of the tridiagonal matrix $T$, that is $\beta_i$ and $\gamma_i$, reach a desired numerical threshold (which is close to zero). However, in practical applications the convergence of the sought after quantities are reached much earlier than that. Therefore, we can truncate the Lanczos chain much earlier at any predefined number. This truncated vector space is called the Krylov subspace. 
Having obtained $\alpha_i$, $\beta_i$ and $\gamma_i$ within the Krylov subspace, we can then evaluate the inverse of the tridiagonal matrix $T$ as a continued fraction. To evaluate CCGF, as it appears in Eq.~\ref{Eq:ccgf2}, we 
evaluate the continued fraction as
\begin{widetext}
\begin{align}
G^{CC}_{pp} (\omega) = \frac{N_{IP}}{(\omega+\mu-i\eta) + \alpha_0 - \frac{\gamma_0 \beta_0}{(\omega+\mu-i\eta) + \alpha_1 - \frac{\gamma_1 \beta_1}{(\omega+\mu-i\eta)+\alpha_2 - ...}}} + 
                      \frac{N_{EA}}{(\omega+\mu+i\eta) - \alpha_0 - \frac{\gamma_0 \beta_0}{(\omega+\mu+i\eta) - \alpha_1 - \frac{\gamma_1 \beta_1}{(\omega + \mu + i\eta) - \alpha_2 - ...}}}. \label{Eq:contfrac}
\end{align}
\end{widetext}
Here, $N_{IP} = \langle \Phi_0 |  (1+ \Lambda) \overline{a_p^\dagger} \overline{a}_p |\Phi_0 \rangle $ and $N_{EA} = \langle \Phi_0 | (1+ \Lambda) \overline{a}_p \overline{a_q^\dagger} | \Phi_0 \rangle $.
Note that in the above expression, the frequency label $\omega$  is completely general enabling evaluation both on the real and imaginary axis by simply changing the definition of the frequency grid.

We should point out here that it is only possible to evaluate diagonal elements of CCGF from Eq. \ref{Eq:contfrac} since only for them the bi-orthonormal condition $P^T Q =1$ is valid.  Lanczos chain vectors are only nearly orthogonal for the off-diagonals elements.
One way to circumvent this problem is to calculate diagonal elements $G_{p+q, p+q}$ instead of off-diagonal elements $G_{pq}$ or $G_{qp}$.
We can then express the $G_{p+q, p+q}$ Green's function in terms individual Green's function elements since in general we can write
\begin{align}
G^{e}_{p+q, p+q} = {} & \langle  \Psi | (a_p + a_q) \frac{1}{\omega-H+E_0} (a_p + a_q)^{\dagger} | \Psi \rangle \nonumber \\
= {} &  G^{e}_{pp} + G^{e}_{qq} + G^{e}_{pq} + G^{e}_{qp}.
\end{align}
This is a general form holding both for the electron and hole Green's functions. 

For a hermitian Hamiltonian, we have in general $G_{pq}=G_{qp}$ and consequently, we can express $G_{pq}=\frac{1}{2}(G_{p+q,p+q}-G_{pp}-G_{qq})$. However, for GFCC the Hamiltonian in Eq. ~\ref{Eq:ccgf2} is non-hermitian, and consequently the Green's function is not symmetric $G_{pq} \neq G_{qp}$. As a result, we can only evaluate $(G_{pq} + G_{qp})=(G_{p+q,p+q}-G_{pp}-G_{qq})$ by Lanczos with the  continuous fraction procedure. 
To evaluate a single off-diagonal element $G_{pq}$, we need to use an approximation and assume symmetrization yielding $G_{pq}=\frac{1}{2}(G_{pq}+G_{qp})$. 
This must be contrasted with the approach we have outlined in Sec. \ref{sec:Lsolver}, where it is possible to calculate whole $G_{pq}$, and not only the symmetric part of it. We have estimated the difference between the ``exact" $G_{pq}$, and only the symmetric part of it. This difference has turned out to be very small. We thus conclude that using this approximation will not have any influence on our results. 

\section{Implementation} \label{sec:implementation}

Here, we describe some aspects of the evaluation of terms involved in Eqs. \ref{Eq:rightLC} and \ref{Eq:leftLC}.

To initiate the Lanczos procedure, we have to first define a set of suitable orthonormal vectors. For the diagonal elements of the h-type Green's function, we choose both right and left hand unit vectors corresponding to the ionization from a specific orbital as guess vectors. The right hand unit vectors are then transformed according to Eq. \ref{Eq:transap}, and left hand unit vectors are transformed as (1+$\Lambda$) $\overline{a}_p^\dagger$. The diagrammatic expressions that arise from all these equations are compiled in Appendix. In the CCSD case, the transformed left and right vectors are up to 2h-1p and 1h-2p rank, respectively for h and e cases. A different situation appears when we calculate the off-diagonal elements of Green's function. In order to evaluate them we have to first calculate $G_{p+q,p+q}$ elements. This requires careful transformation of $a_p$ and $a_p^\dagger$ operators. We have explained this aspect in Appendix.

The dominant cost in this procedure is the evaluation of matrix-vector products $\overline{H}q_i$ and $p_i^T\overline{H}$. These quantities are evaluated exactly in the same way as in the right and left hand side of the equation of motion ionization potential (EOMIP) and the equation of motion electron affinity (EOMEA) $\sigma$-vector equations. Thus, the cost of these operations is $n^5$, where $n$ is the number of MO basis functions.   
\begin{figure} [htb]
\includegraphics[width=\columnwidth]{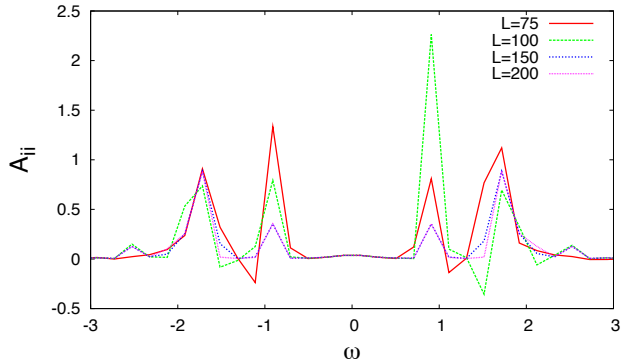} \caption{Convergence of a spectral function with respect to the number of Lanczos vectors. L stands for the number of Lanczos vectors.}
\label{fig:Leg_convergence}
\end{figure}

\section{Results and Discussion} \label{Sec:results}

The aim of this paper is to investigate the applicability of the GFCC method as an impurity solver. Impurity problems appear frequently as auxiliary models in condensed matter physics~\cite{Georges96} and more recently in quantum chemistry where they are an integral part of the construction for embedding methods~\cite{ZgidPRB2012, Zgid11}. Consequently, good solvers capable of yielding a Green's function and self-energy are desired. 
Most of new solvers are tested on impurity problems arising in the treatment of the 1D and 2D Hubbard models parametrized with various interaction strengths.  These test cases provide an estimation of applicability of such solvers to specific materials and can be easily compared to other established computational methods. 

The Hubbard model is defined by the following Hamiltonian 
\begin{equation}
H = - \sum_{<ij>,\sigma} t_{ij} a^\dagger_{i\sigma} a_{j\sigma} + \sum_{i, \sigma \neq \sigma'} U_{ii} n_{i\sigma}n_{i\sigma'},  
\end{equation}
where, $<ij>$ stands for a summation over the nearest neighbors, $t$ is the hopping integral between nearest neighbor sites $i$ and site $j$, and $U>$ 0 is the on-site interaction. The occupation number at site $i$ for spin $\sigma$ is defined as $n_{i\sigma} = a_{i\sigma}^\dagger a_{i\sigma}$. 

In order to produce the Anderson impurity model (AIM) described by a Hamiltonian 
\begin{align}
H_{imp}=H_{cluster}+H_{bath}+H_{int},
\end{align}
where the respective parts are: the cluster Hamiltonian $H_{cluster}$, the bath Hamiltonian $H_{bath}$, and the Hamiltonian of the cluster-bath interaction $H_{int}$ that are expressed as
\begin{align}\label{eq:impurity_ham}
H_{cluster} ={} & - \sum_{<uv> \in cluster,\sigma} t_{uv} a^\dagger_{u\sigma} a_{v\sigma}+ U \sum_{u \in cluster} n_{u\uparrow}n_{u\downarrow}, \nonumber\\
H_{bath} =  {} &  \sum_{b \in bath,\sigma} \epsilon_b a_{b\sigma}^\dagger a_{b\sigma}, \nonumber\\
H_{int}={} & \sum_{u \in cluster, b \in bath, \sigma} V_{ub\sigma} (a_{u\sigma}^\dagger a_{b\sigma} + a_{b\sigma}^\dagger a_{u\sigma}),
\end{align}
 we perform a projection from the Hubbard lattice to the impurity problem by demanding that the local part of the lattice Green's function is the same as the impurity Green's function
\begin{equation}
[G_{lattice}(\omega_n)]_{local}=G_{imp}(\omega_n).
\end{equation}
Here, the impurity Green's function is defined as  
\begin{equation}
G_{imp}(\omega_n)=[\omega_n+\mu -t-\Sigma(\omega_n)-\Delta(\omega_n)]^{-1}\label{Eq:gimp}.
\end{equation}
The hybridization of the impurity orbital with other orbitals present in the problem is denoted as $\Delta(\omega_n)$. Using a finite bath containing $n$ sites it is expressed as 
\begin{equation}
\Delta_{uv} (\omega_n) \approx \sum_b^n \frac{V^*_{ub} V_{vb}}{\omega_n - \epsilon_b}\label{Eq:hybridization}.
\end{equation}

In order to investigate the applicability of CCGF as an impurity solver, we test it on impurity problems with different interaction strength $U/t$ for both the 1D and 2D Hubbard models. After defining an impurity Hamiltonian as listed in Eq.~\ref{eq:impurity_ham} and defining the filling (e.g. half-filling), that influences the number of electrons present in the impurity model, we carry out the following computational steps:
\begin{enumerate}
\item Since the CC method used here is a zero temperature method, it is necessary to find a ground state of the impurity problem. This is done by searching for the lowest energy among all particle numbers $N$ and $S_z$ sectors possible for a given impurity.  
\item Carry out CCSD calculation to obtain T1 and T2 amplitudes on the impurity problem with $(N,S_z)$ yielding the lowest energy.
\item Solve CCSD left-eigenvector problem to extract de-excitation amplitudes $\Lambda_1$ and $\Lambda_2$.
\item Evaluate all hole and particle type Green's function elements, G$_{pq}^{(h)}$ and G$_{pq}^{(e)}$, and sum them up to obtain total Green's function $G_{pq}$. We then transform that Green's function to the site basis to obtain only the impurity Green's function, $[G_{imp}]_{uv}$.
\item Evaluate the impurity self-energy using the Dyson's equation
\begin{equation}\label{eq:dyson}
\Sigma_{imp}(\omega_n) = [G_0(\omega_n)]^{-1} - [G_{imp}(\omega_n)]^{-1}. 
\end{equation}
Here, $G_{0}$ is the non-interacting Green's function evaluated as
\begin{equation}
[G_{0} (\omega_n)]_{uv} = [\omega_n+\mu - t_{uv} - \Delta_{uv}(\omega_n)]^{-1} \label{Eq:bareG},
\end{equation}
where, the definition of $\Delta(\omega_n)$ is in Eq. \ref{Eq:hybridization}.   
\end{enumerate}


\subsection{1D Hubbard}\label{sec:1D Hubbard}
In this work for both 1D and 2D Hubbard model the energy and value of the on-site integral $U$ will be expressed in the units of hopping integral $t$, where $t=1$.
The first model we have considered is the 1D Hubbard model, where after the projection to an impurity model, the resulting Anderson impurity model is described by a single impurity orbital coupled to a bath. 
For the value of inverse temperature $\beta=400/t$, we achieved a good fit to Eq.~\ref{Eq:hybridization} with 11 bath orbitals. At the half-filling, there are  12 electrons in the system, and we first converge to the (6$\uparrow$, 6$\downarrow$) closed-shell configuration at the HF level. The ensuing CC calculation is carried out on that reference. 

\subsubsection{Convergence with respect to the number of Lanczos vectors}

First, for the 1D Hubbard model, we analyze the convergence of the spectral function defined as 
\begin{equation}\label{eq:spectralf}
A(\omega)=-\frac{1}{\pi}{\rm Tr}({\rm Im}[G(\omega)] )    
\end{equation}
as a function of the number of Lanczos vectors. From Fig. \ref{fig:Leg_convergence}, we observe that the evaluation of an accurate spectral function requires a large number of Lanczos vectors. In this particular case, for small numbers of Lanczos vectors $75 \le L <150$, the spectral function is inaccurate and has negative values. Only for $L\ge 150$, we obtain a converged positive spectral function.   

\subsubsection{Self-energy for different values of $U/t$ on the imaginary axis}

Here, for different values of $U/t$, we compare the GFCC self-energies to the ones coming form FCI or its truncated versions. Comparing self-energies instead of Green's functions is a particularly stringent test of accuracy since it highlights the direct differences in the computed quantities. 
For the particle-hole symmetric case, we only plot the imaginary part of self-energy since the real part is zero. We have compared  GFCCSD with FCI and two truncated CI versions, namely, CISD and CISDT. For values of U/t=4, 6 and 8, we observe from Fig. \ref{fig:1DSigma} that CISD performs poorer at low frequency regime than CISDT or CCSD.  The behavior of both CISD and CISDT worsens as we increase the interaction strength. CCSD agrees almost exactly with FCI. The only deviation present is for the lowest frequencies where the self-energy is very close to zero.
\begingroup
\centering
\begin{figure*}[]
\centering
\subfigure{\includegraphics[width=0.32\textwidth]{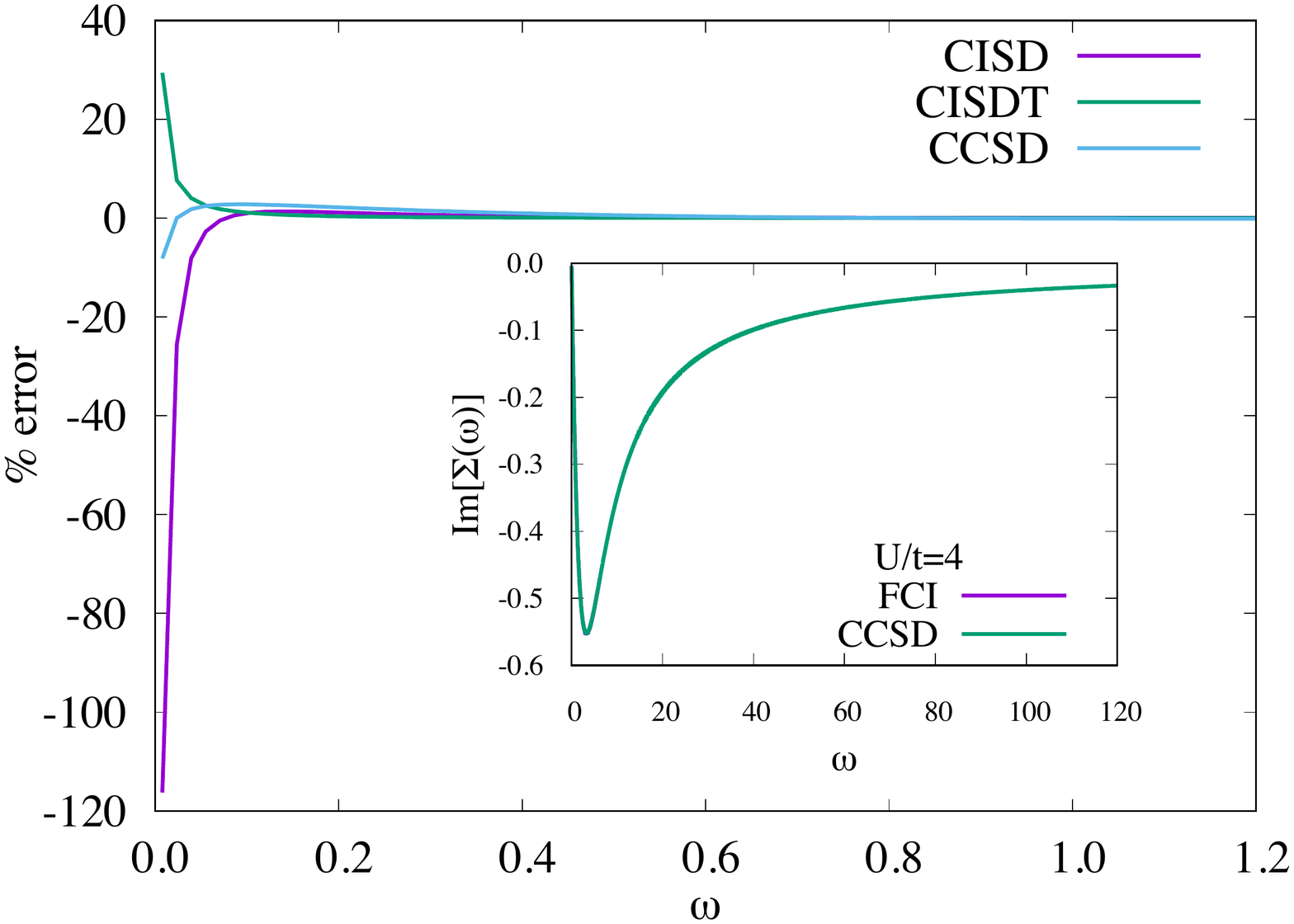}}\label{fig:1a}
\subfigure{\includegraphics[width=0.32\textwidth]{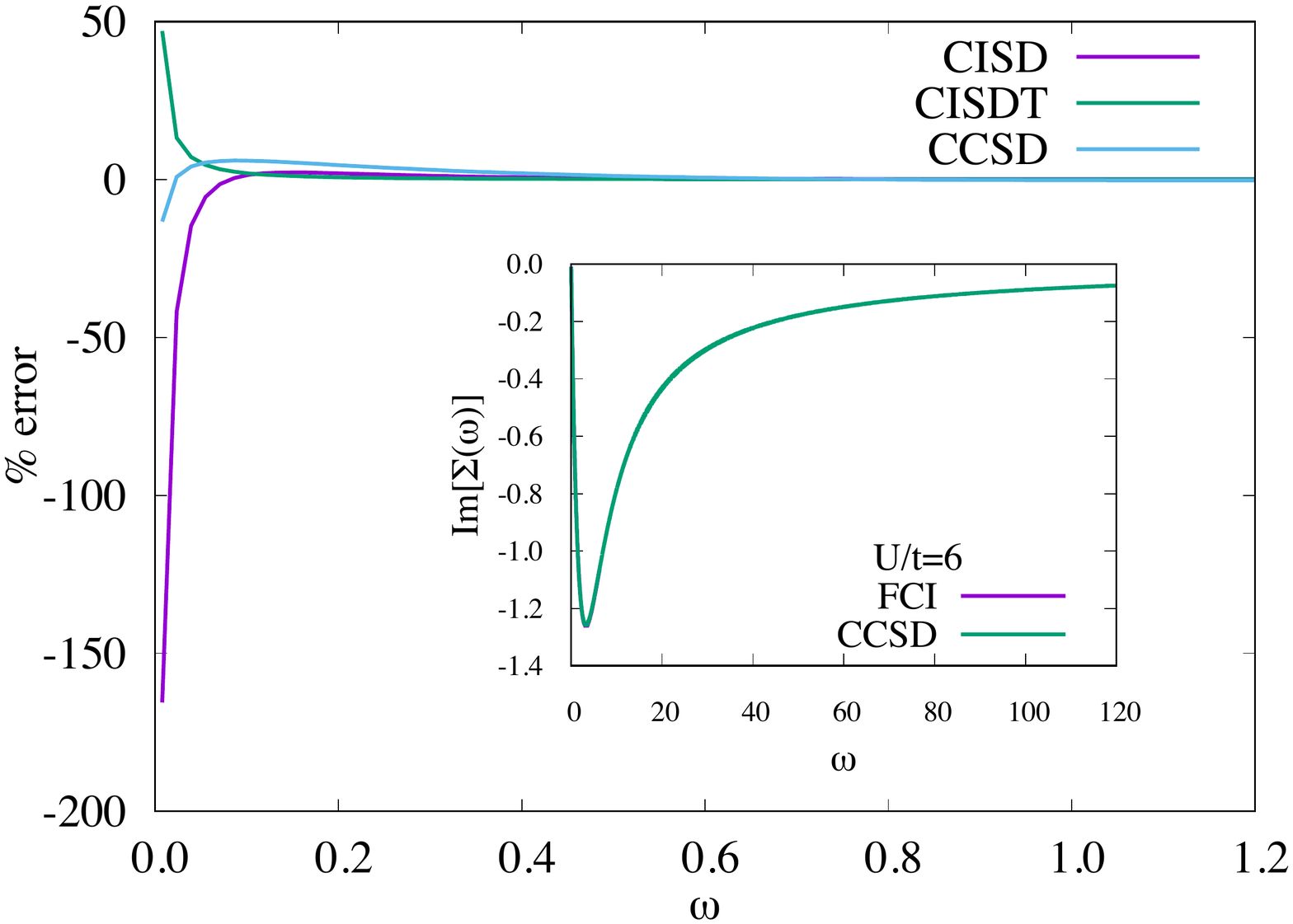}}\label{fig:1b}
\subfigure{\includegraphics[width=0.32\textwidth]{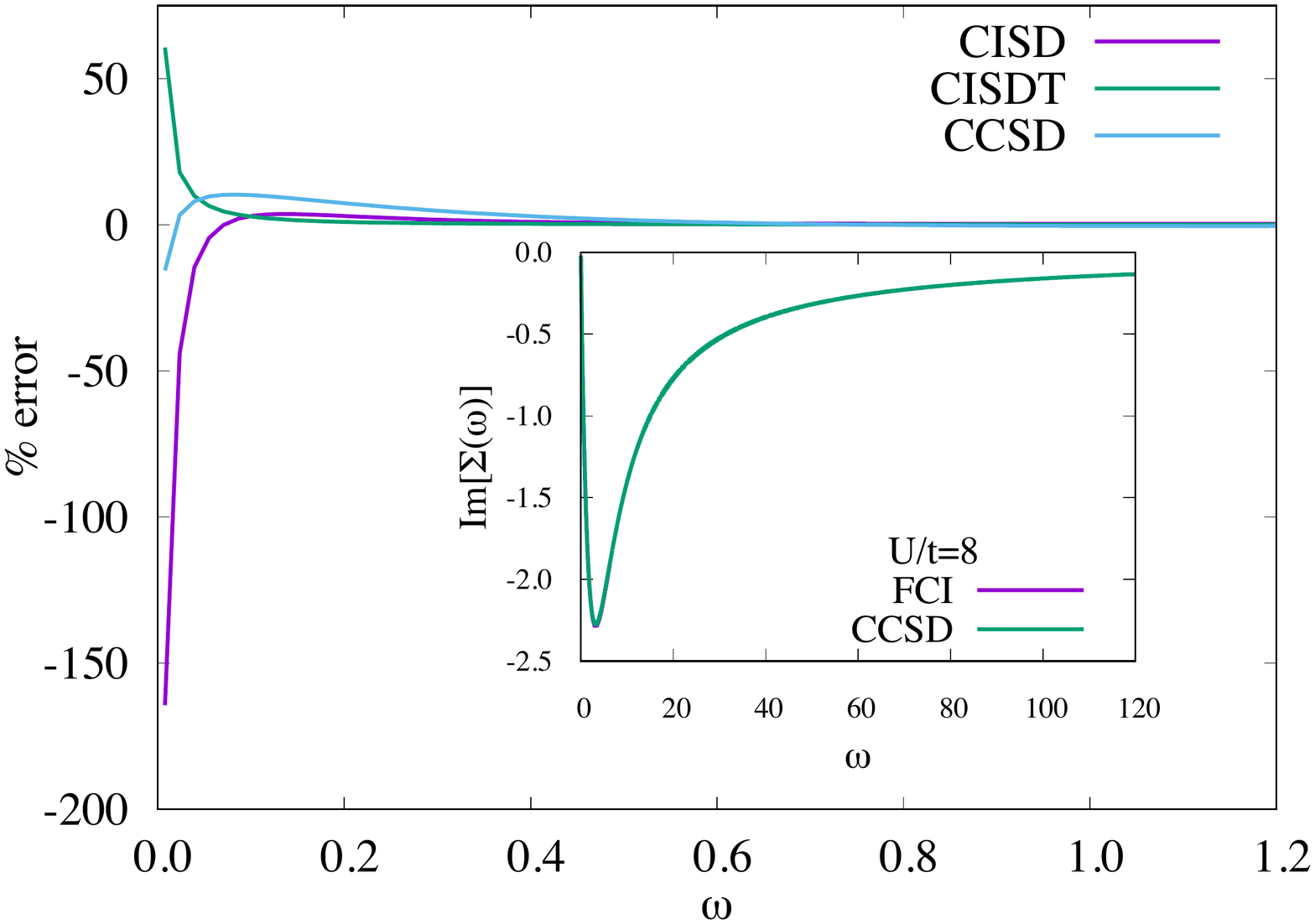}}\label{fig:1c}
\caption{The errors of CISD, CISDT, and CCSD self-energies in comparison to the FCI one for 1D Hubbard model with different interaction strengths U/t. The insets show the FCI and CCSD self-energy. All the calculations were performed for the inverse temperature $\beta t=400$.} \label{fig:1DSigma}
\end{figure*}
\endgroup

\subsubsection{Spectral functions}
We have compared spectral functions from GFCC at the CCSD level with spectral functions obtained using FCI as well as its truncated versions, here particularly CISD and CISDT. In Fig.~\ref{fig:1DSpectra}, for all ranges of $U/t$, we observe an excellent agreement between spectral functions coming from CCSD, FCI, and its truncated version. The spiky character of the spectral functions is caused by a small number of bath orbitals. 
\begingroup
\centering
\begin{figure*}[]
\centering
\subfigure{\includegraphics[width=0.32\textwidth]{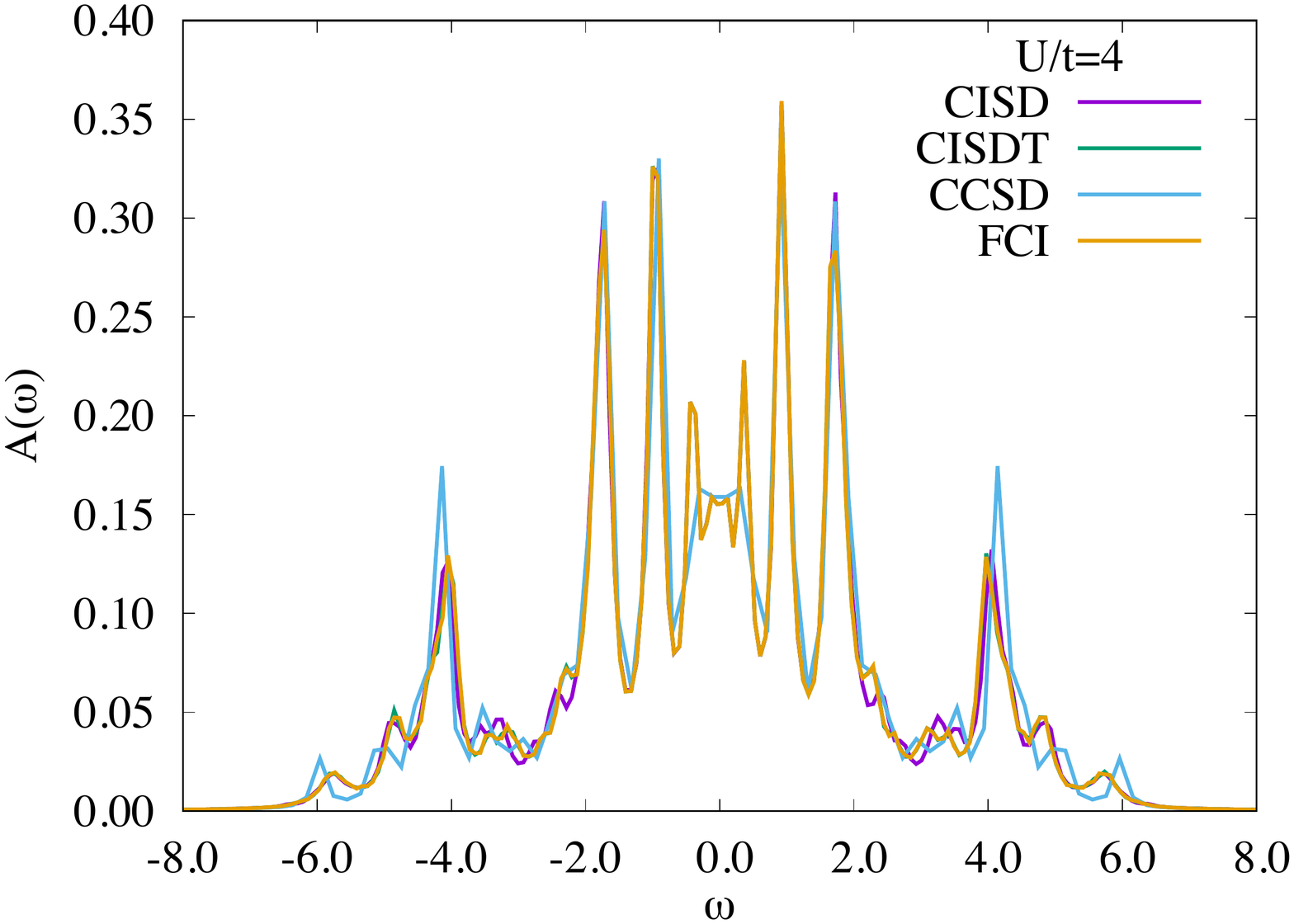}}
\subfigure{\includegraphics[width=0.32\textwidth]{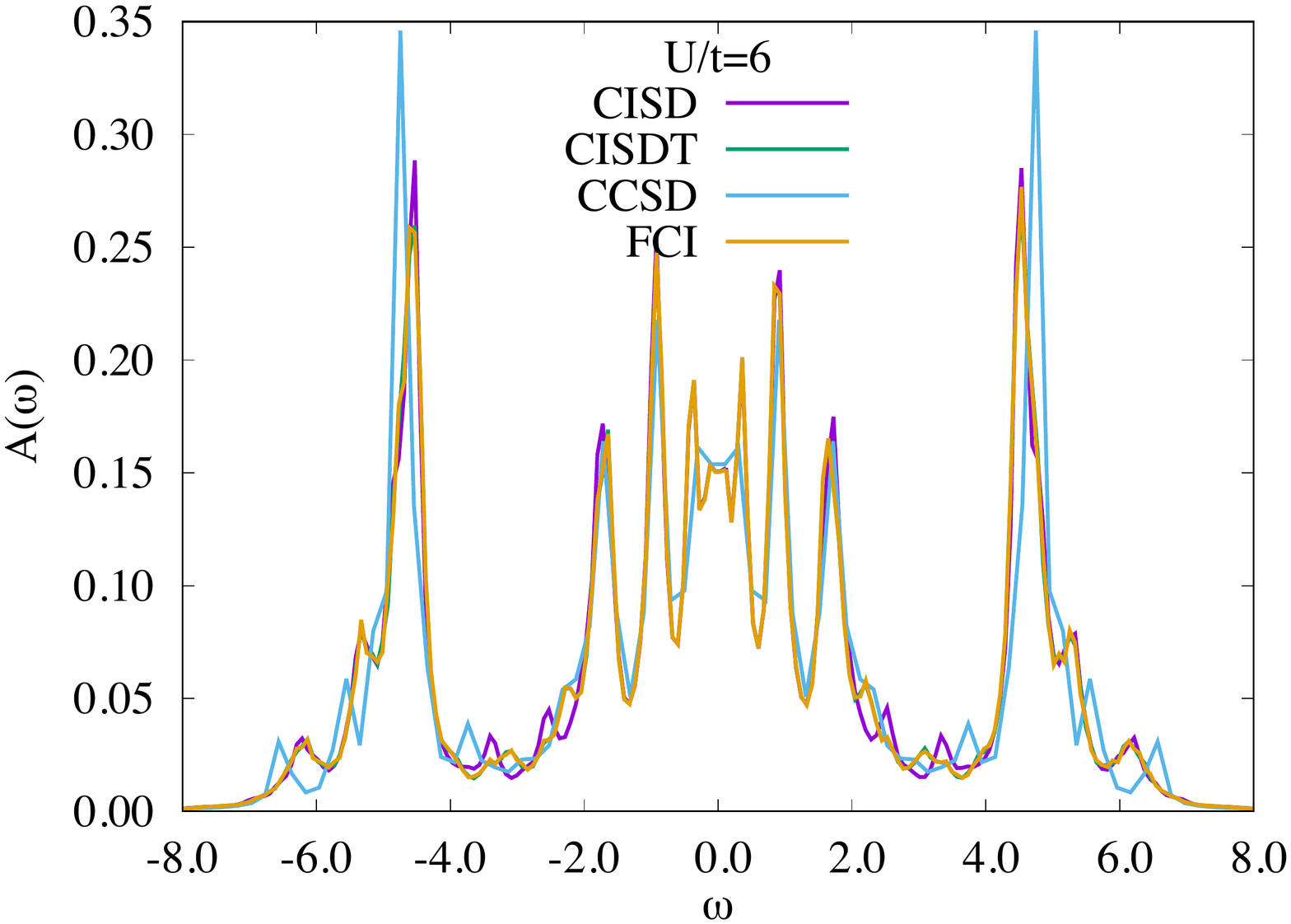}}
\subfigure{\includegraphics[width=0.32\textwidth]{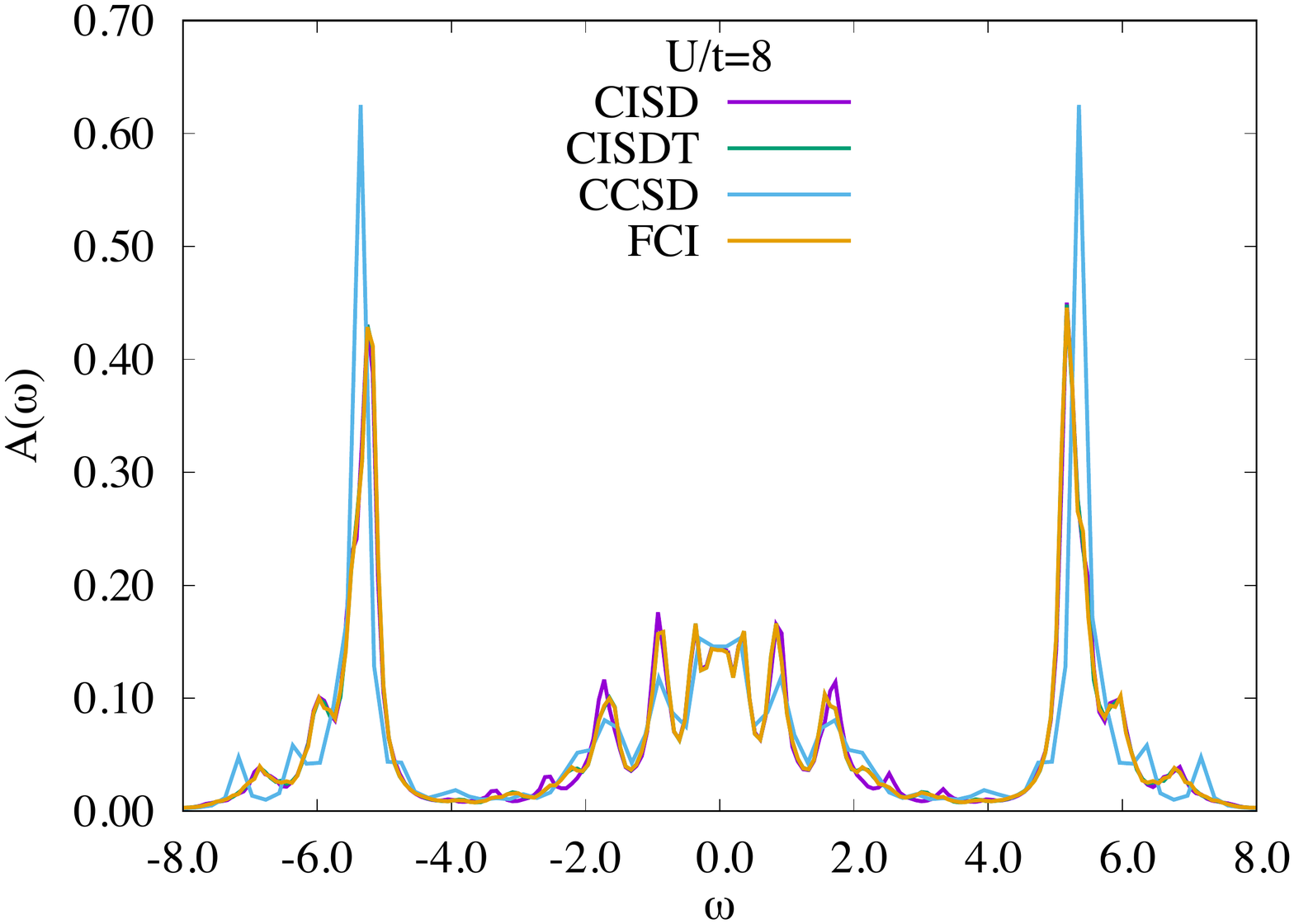}}
\caption{Spectral functions for 1D Hubbard model from CISD, CISDT, FCI, and CCSD  with different interaction strengths $U/t$. All the calculations were performed for the inverse temperature $\beta t=400$.} \label{fig:1DSpectra}
\end{figure*}
\endgroup

\subsubsection{Convergence with respect to the number of bath orbitals}

\begin{figure} [htb]
\includegraphics[width=\columnwidth]{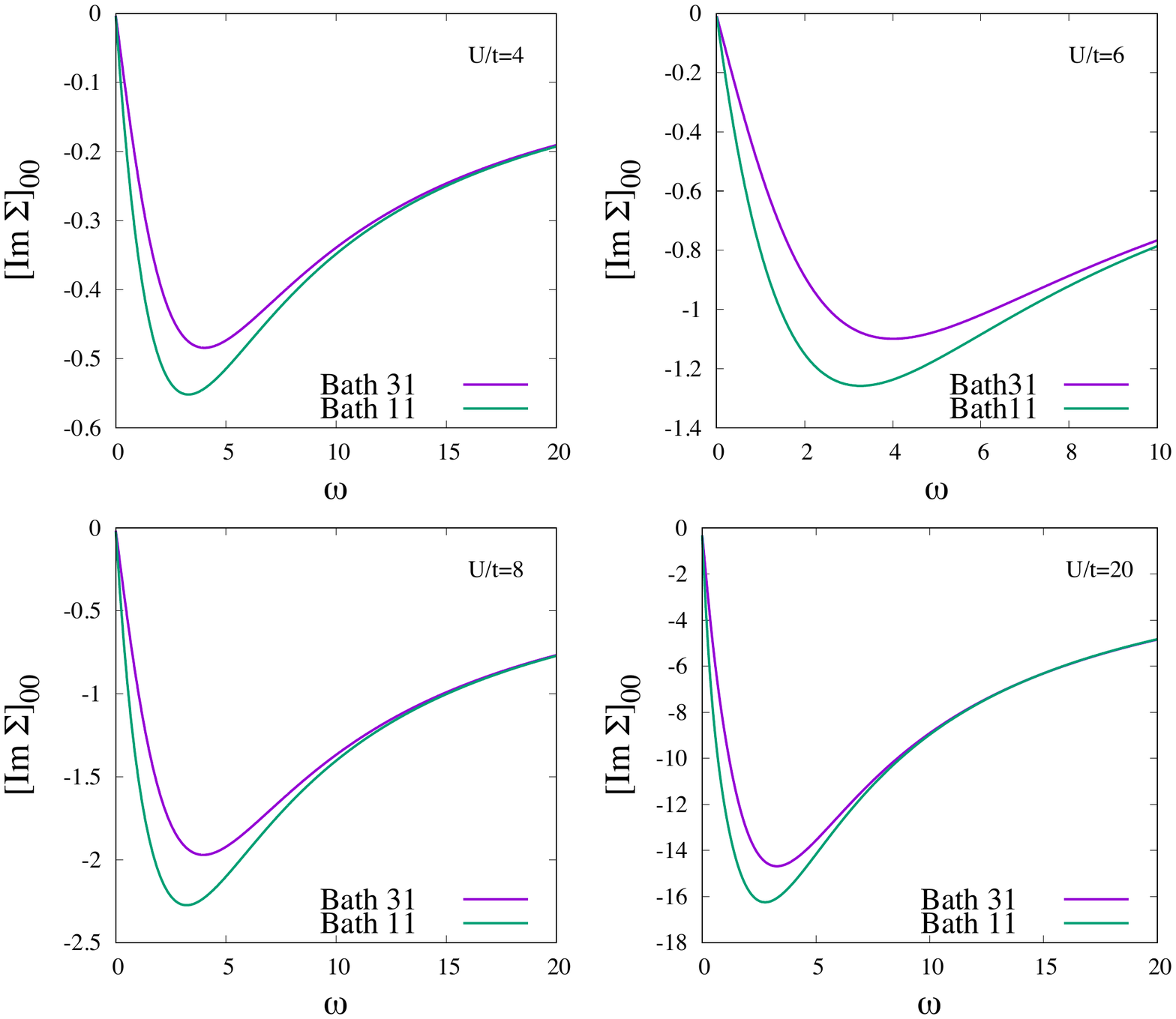} \caption{Imaginary self-energy plots for 1D Hubbard model at the CCSD level with varying number of bath orbitals.}
\label{fig:bath}
\end{figure}

We investigated the convergence of the self-energy with respect to the number of orbitals used to fit the hybridization function. 
Since an evaluation of GFCC at the CCSD level has much more preferable scaling $O(n^6)$, where $n$ is the number of orbitals, to the exponentially scaling FCI, one can assume that calculations with larger number of bath orbitals and/or for larger number of impurity orbitals are possible.
Consequently, in our calculations, we have increased the number of bath orbitals to investigate how much self-energy differs with that choice. 
Here as an illustration, we have considered a single impurity coupled to 11 and 31 bath orbitals. Resulting imaginary parts of self-energy are  plotted in Fig.~\ref{fig:bath}. 
While both self-energies at the qualitative level do not seem to differ, the quantitative differences between them seem indicate that GFCC due to the possibility of treating many bath orbitals will be very useful for analyzing the convergence with respect to the number of bath orbitals.

\subsection{2D Hubbard}\label{sec:2d_hubbard}
The 2D Hubbard lattice was mapped onto an impurity model containing a $2\times 2$ impurity cluster with 8 bath orbitals (2 bath orbitals per impurity site). Note, that while for the 1D Hubbard model, we only carried out restricted CC calculations, here for the 2D Hubbard model, we use unrestricted CC. We will consider here, only the 2D Hubbard model at half-filling since based on our previous experience the regimes away from half-filling were usually simpler for quantum chemical type solvers~\cite{ZgidPRB2012,Zgid11}.

\subsubsection{Determination of the reference state for unrestricted coupled cluster}

\begin{figure} [htb]
\includegraphics[width=\columnwidth]{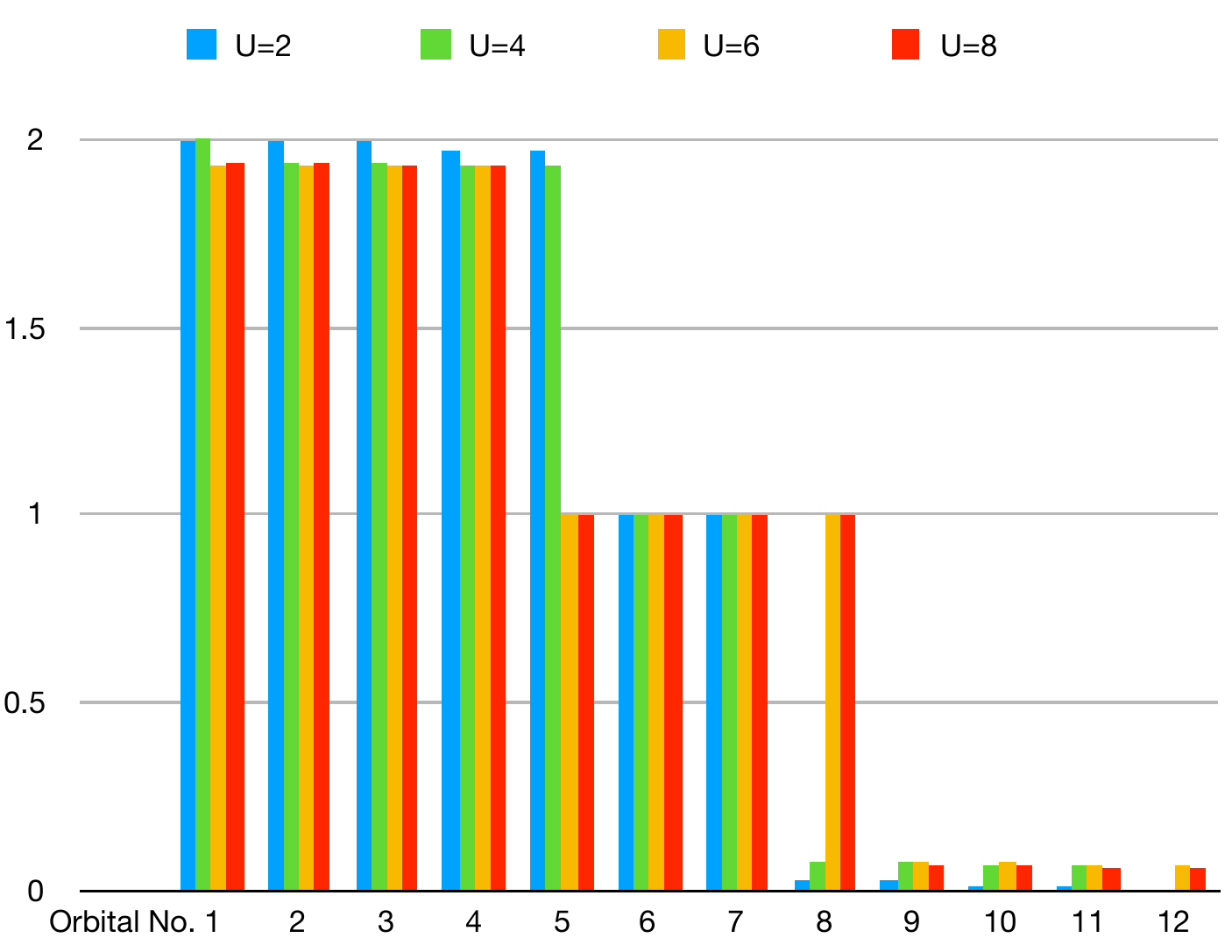} \caption{Orbital occupancies of the lowest-energy state from FCI.}
\label{fig:fci_occ}
\end{figure}

For an impurity model defined in such a manner,  we have carried out FCI calculations to determine the lowest energy wavefunction and the occupation of orbitals. In Fig.~\ref{fig:fci_occ}, we list occupancies of the impurity  orbitals obtained from FCI for different $U/t$ regimes. 

Subsequently, in order to carry out open-shell CC calculations, we have chosen reference determinants for the unrestricted CC calculations based on the determinantal expansion of the FCI wavefunction. 
 To start the CC calculations, we have chosen the highest multiplicity determinant from a number of open-shell determinants that were present in the FCI expansion. Although the resulting CC wavefunction suffers from a spin-contamination error, the resulting energies give us information about the performance of CC for these states. 
The deviation from FCI  of the total CC energy with different levels of amplitude truncation is shown in Table.~\ref{Tab:2DHub_tot}.

At $U/t=2$ and $U/t=4$, we obtain the lowest FCI energy for a wave function with two dominant open-shell determinants. In both of these regimes, energies evaluated within UCCSD behave very well and the CC energy improves only very slightly with the increase of the excitation level.

 In contrast, for $U/t=6$ and $U/t=8$ regimes, the FCI wave functions were dominated by four open-shell determinants and therefore for these regimes, we observe a large deviation of the CC energies from FCI and a lack of monotonous convergence to the FCI energy when the excitation level is increased.
\begin{table}
\caption{Comparison of total energies as calculated from CC with different excitation level against FCI. $\Delta$E$_\text{X}$ stands for E$_\text{X}$-E$_{\rm FCI}$. ``Reference'' denotes high spin that were used for CC calculations.} 
\begin{tabular}{|c|c|c|c|c|c|}
\hline 
U/t & Reference & $\Delta$E\textsubscript{CCSD} & $\Delta$E\textsubscript{CCSDT} & $\Delta$E\textsubscript{CCSDTQ} & E\textsubscript{FCI}\tabularnewline
\hline 
\hline 
2 & 7$\alpha$5$\beta$& 0.0039 & 0.0038 & 0.0038 & -12.598019\tabularnewline
\hline 
4 & 7$\alpha$5$\beta$ & 0.0056 & 0.0043 & 0.0039 & -14.947579\tabularnewline
\hline 
6 & 8$\alpha$4$\beta$& 0.1887 & 0.1875 & 0.1877 & -17.848516\tabularnewline
\hline 
8 & 8$\alpha$4$\beta$& 0.0642 & 0.0640 & 0.0640 & -21.329302\tabularnewline
\hline 
\end{tabular} \label{Tab:2DHub_tot}
\end{table}

In practical calculations for impurity problems, we do not know the number of electrons that is present in the impurity. Therefore, as described earlier multiple calculations with different number of electrons and different starting open-shell determinants will need to be performed to find the lowest energy solution.
This search has be executed with caution since CC 
has no lower bound on energy, thus cases where divergences happen may yield very low energies. However, we believe that analyzing the behavior of CC with a series of excitation together with amplitude analysis can guide users to finding a proper number of electrons within the impurity. 

\subsubsection{Comparison of self-energies for 2D Hubbard at half-filling}

\begin{figure*} [htb]
\includegraphics[width=\columnwidth]{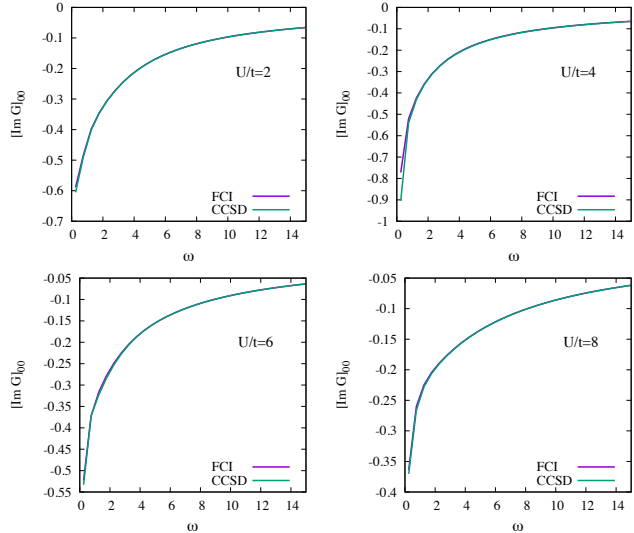} \caption{Comparison between CCSD and FCI Green's function for 2D Hubbard model at half-filling for different interaction strengths. Note that FCI and CCSD Green's functions differ for small frequencies for $U/t=6$ and $U/t=8$. We used $6\alpha6\beta$ state to evaluate the CCSD Green's function.}
\label{fig:2DSigma_G}
\end{figure*}

\begin{figure*} [htb]
\includegraphics[width=\columnwidth]{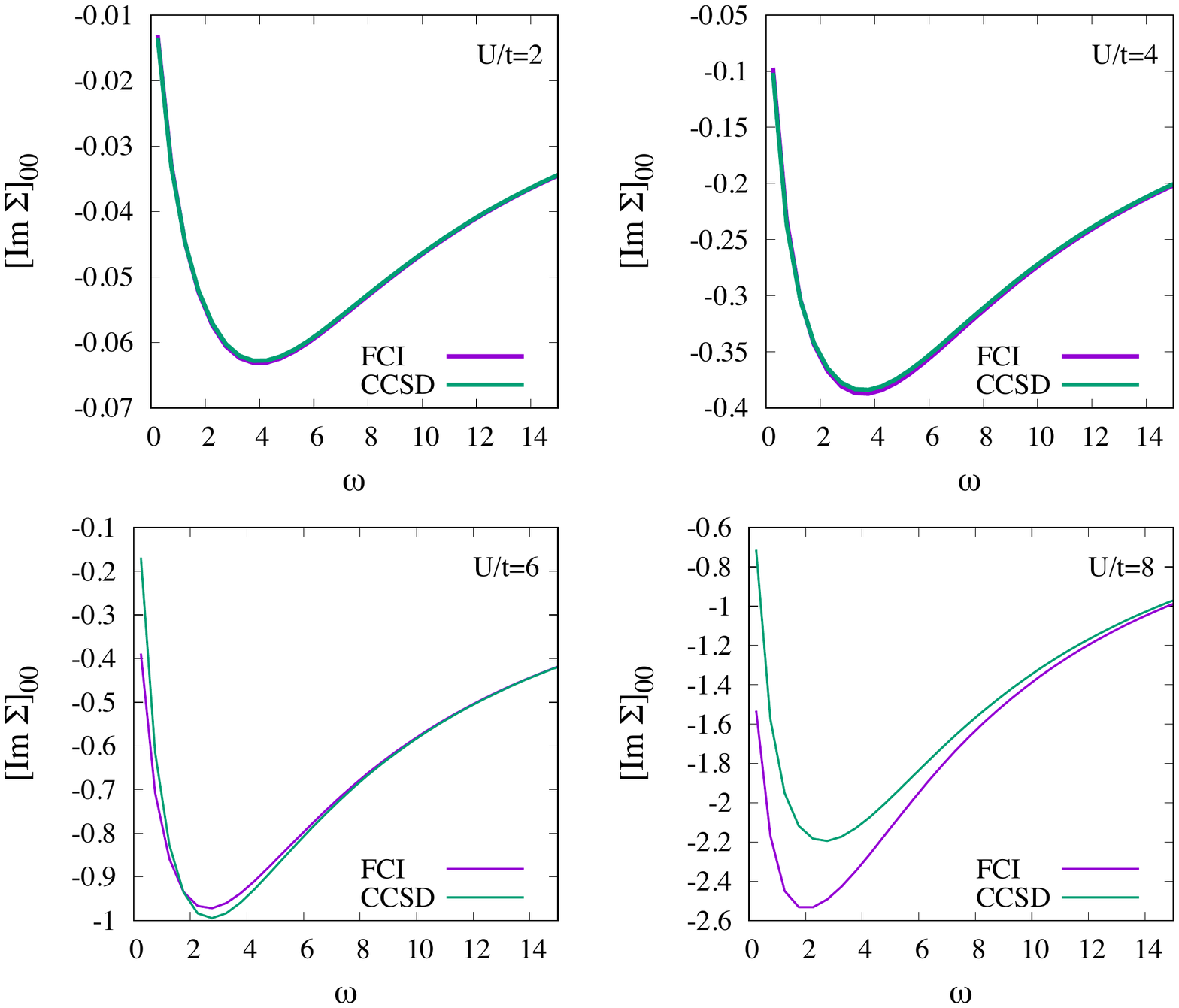} \caption{Comparison between CCSD and FCI self-energies for 2D Hubbard model at half-filling for different interaction strengths.}
\label{fig:2DSigma_sigma}
\end{figure*}

In Fig. \ref{fig:2DSigma_G}, we are comparing the $[G(i\omega)]_{11}$ element of the CCSD Green's function with FCI. Note that we have employed an orbital basis that diagonalizes the $2\times2$ impurity Green's function. For details of how to construct such a basis see Ref.~\onlinecite{ZgidPRB2012}. The behavior for other elements of Green's function follows a similar trend. In order to evaluate Green's function, we have calculated both G$_{\alpha \alpha}$ and G$_{\beta \beta}$ components of the Green's function and then estimated the total Green's function as $G=\frac{1}{2} (G_{\alpha \alpha} + G_{\beta \beta})$. We observe that the CCSD Green's function $[G(i\omega)]_{11}$ element is in a  very good agreement with FCI for the $U/t=2$ and $U/t=4$ regimes for all frequency ranges. For $U/t=6$ and $U/t=8$ there is a  deviation at the lowest frequency points. Since comparing Green's functions alone can hide important differences, we also have calculated the self-energy using Eq.~\ref{eq:dyson} as described in Sec.~\ref{Sec:results}. The $[\Sigma(i\omega)]_{11}$ element is plotted in comparison to FCI in Fig.~\ref{fig:2DSigma_sigma}. While for the weakly correlated regimes, $U/t=2$ and $U/t=4$, the CCSD self-energies are almost exact, we observe an increasing deviation from the exact result at low and intermediate frequency range for $U/t=6$ and $U/t=8$. This result is expected since for both of the more strongly correlated regimes, we observe an increasing number of unpaired electrons that cannot be described exactly at the CCSD level. In Tab.~\ref{tab:ccsd_2d_hub_occupancy} we list impurity orbital occupancies for different $U/t$ regimes.  We surmise that evaluating Green's function and subsequently self-energy at the CCSD(T) or CCSDT level should minimize the differences with respect to FCI. 
\begin{table} [htb]
\caption{Impurity orbital occupation numbers for 2D Hubbard model at half-filling obtained at the CCSD level.}
\begin{tabular}{lllll}
Orbital No. & U/t=2 & U/t=4 & U/t=6 & U/t=8\\
\hline 
1 & 1.9999 & 1.9831 & 1.9654 & 1.9715 \\
2 & 1.9912 & 1.9598 & 1.9654 & 1.9715 \\                            
3 & 1.9911 & 1.9497 & 1.9369 & 1.9371 \\
4 & 1.9743 & 1.8722 & 1.9327 & 1.9334 \\
5 & 1.9729 & 1.8436 & 1.0087 & 1.0067 \\
6 & 0.9999 & 1.0335 & 0.9999 & 1.0000 \\
7 & 0.9999 & 0.9531 & 0.9999 & 1.0000 \\
8 & 0.0267 & 0.1742 & 0.9904 & 0.9929 \\
9 & 0.0264 & 0.1221 & 0.0725 & 0.0701 \\
10 & 0.0088 & 0.0480 & 0.0584 & 0.0599 \\
11 & 0.0087 & 0.0435 & 0.0350 & 0.0285 \\
12 & 0.0001 & 0.0098 & 0.0346 & 0.0284 \\
\end{tabular}\label{tab:ccsd_2d_hub_occupancy}
\end{table}
\subsection{One-body density matrix and total energy}

Many Green's function methods (e.g. Green's function second order (GF2) or fully iterative GW schemes) are based on the diagrammatic expansion of the Luttinger-Ward $\Phi$-functional~\cite{Luttinger60}. Here, the self-energy is a functional derivative of the $\Phi$-functional over Green's function. These methods are conserving and phi-derivable. This means that quantities such as  electronic energy that are evaluated in these method will give the same results independent of the way these quantities are evaluated. For example, the electronic energy can be evaluated from thermodynamic integration or Galitskii-Migdal formula yielding exactly the same result. This is not the case for the CC Green's function since the CC method is not written as a diagrammatic expansion of a $\Phi$-functional and a diagrammatic expansion is only given for a Green's function alone. Consequently, CC self-energy is not a functional derivative of the $\Phi$-functional. It has to be evaluated as described in Sec.~\ref{Sec:results} using Eq.~\ref{eq:dyson}.
Consequently, CC methods are not phi-derivable. This means that the CC electronic energy when evaluated from the CC equations can differ from the same quantities that are evaluated using both CC Green's functions and self-energies.

 The one-body density matrix is evaluated from the imaginary Green's function as 
\begin{equation}
\gamma_{pq} = - \sum_\sigma G_{pq, \sigma} (\tau = \beta), \label{Eq:density}
\end{equation}
where $\sigma$ is denoting a spin label.


To evaluate the electronic energy we use the Galitskii-Migdal formula~\cite{Galitskii1958} 

\begin{align}
E^{2b}_{GM} &= \frac{1}{\beta} \sum_{pq,\sigma} \sum_\omega G_{pq,\sigma} (\omega) \Sigma_{pq, \sigma} (\omega),\\
E^{1b}_{GM}&=\frac{1}{2} Tr [(h^{AO} + F^{AO}) \gamma^{AO} ],\\
E&=E^{1b}_{GM}+E^{2b}_{GM},
\label{Eq:GM-formula}
\end{align}
where we use a Green's function that produces the correct number of electrons for the system and we calculate the self-energy via the Dyson equation. 
The one-body density matrix $\gamma^{AO}$ is evaluated from the CC Green's function using Eq.~\ref{Eq:density}.

In this section, we present a comparison between the CCSD one-body density matrix and electronic energy as we recover them from GFCC  with the ones obtained directly from the CC parent calculations.
We have studied several small molecular systems - H$_2$O, H$_2$S, NH$_3$, PH$_3$, HF, NF$_3$ to analyze the differences. 

Since the imaginary axis GFCC is evaluated on a grid with a spacing of $\beta$, we have chosen the value of $\beta$ such that the calculations are well converged with the spacing. Additionally, we have chosen a high number of frequency grid points to  converge the electronic energy with respect to the grid. These converged grids allow us to compare the energies obtained from the Green's function CC and parent CC equations. 
The detailed comparison of energies as well as occupation numbers is presented in Tab.~\ref{tab:gfcc_energy}. The occupation numbers and the density differ only very slightly from the parent CC result since density matrix evaluated in CC and through GFCC should be the same ~\cite{NooijenIJQC1992}. The differences present here can be attributed to the inexactness of the grid. However, the differences in electronic energies are not just a mere artifact of the grid. They truly come from the fact that the self-energy that is produced by a CCGF is not a functional derivative of the $\Phi$-functional with respect to the Green's function. Consequently, when a CCGF is used in an embedding calculations a special care must be taken to compare energies with other calculations. They can only be compared with the ones that were evaluated exactly in the same way. 
Since in embedding calculations, we will necessarily evaluate energy using the Galitskii-Migdal formula, we need to compare this energy to other energies also evaluated using this formula.

\begin{table} [htb]
\caption{Energy and density evaluated from CCSD Green's function. MUE $\equiv$ Mean Unsigned Error ; MSE $\equiv$ Maximum Signed Error ; $\Delta$ E$_{CC}$ = E$_{CCSD}$ - E$_{CCSD}^{GM}$}
\begin{tabular}{|c|c|c|cc|c|}
\hline 
 &  & \multicolumn{1}{c}{} & Occ. No. &  & Energy\\
\hline 
\hline 
Molecules & Basis & MUE & MSE &  & $\Delta$ E$_{CC}$\\
\hline 
H$_2$S & STO-6G & 0.00005& 0.0002 &  & -0.008\\
\cline{2-6} 
 & cc-pVDZ & .00005 & 0.0003 &  & -0.012\\
\hline 
NH$_3$ & STO-6G & 0.00004 & 0.0001 &  & -0.0067 \\
\cline{2-6} 
 & cc-pVDZ & 0.00002 & 0.0001 &  & -0.0191  \\
\hline 
PH$_3$ & STO-6G & 0.00005 & 0.0002 &  & -0.0047 \\
\cline{2-6} 
 & cc-pVDZ & 0.00009 & 0.0005 &  & -0.0171\\
\hline 
HF & STO-6G & 0.00003 & 0.0001 &  & 0.0111\\
\cline{2-6} 
 & cc-pVDZ & 0.0005 & 0.0037 &  & -0.0553\\
\hline 
NF$_3$ & STO-6G & 0.00038 & 0.0014 &  & -.042\\
\cline{2-6} 
 & cc-pVDZ & 0.0003 & 0.0028 &  & -0.0035\\
\hline 
\end{tabular}\label{tab:gfcc_energy}
\end{table}

\subsection{Self Energy Embedding Theory with CCSD as a Solver}\label{sec:seet_gfcc}

In this section, we explore a possibility of using the CCSD self-energy in a Green's function based embedding framework, namely the self energy embedding theory (SEET)\cite{SEET16, zgid_njp17}.  Originally, SEET was created to describe strongly correlated molecular problems. In SEET, the whole system of interest was separated into strongly correlated orbitals embedded in the weakly correlated environment (for details see Ref.~\onlinecite{SEET16,zgid_njp17}). The weakly correlated environment was then treated with a  low-level, most frequently perturbative method, whereas the strongly correlated orbitals were handled  with a high-level, usually exponentially scaling, non-perturbative  method. Depending on the complexity of the problem various high-level correlated methods can be chosen in this framework.  

The results that we obtained for 2D Hubbard model in Sec.~\ref{sec:2d_hubbard} suggest that Green's function CCSD solver is better suited to handle weakly correlated problems.
Consequently, here, we do not aim to solve any strongly correlated problems and we will be using the embedding framework only to separate the orbitals into orbital groups that require different levels of treatment. We only assume that in the embedding construction there are orbital groups that require more accurate treatment than the  environment that will be described at a more approximate level. Using such a definition of the embedding problem, we aim to describe accurately large weakly correlated problems.

\begin{figure*} [htb]
\includegraphics[width=\columnwidth]{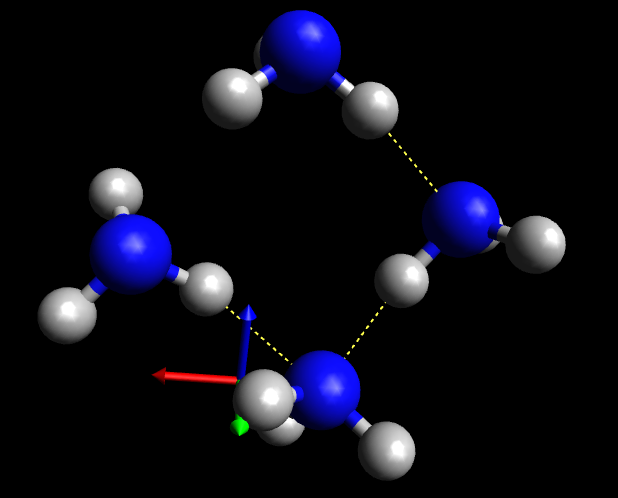} \caption{Structure of ammonia tetramer - blue spheres represent N, white spheres H and yellow dotted lines are for hydrogen bonding.}
\label{fig:tetramer}
\end{figure*}

One prototypical group of weakly correlated problems that require an accurate treatment are molecular crystals with non-covalent bonding. Thus, to check the performance of the embedding scheme, we have chosen a model system such as a tetramer of ammonia as in Fig. \ref{fig:tetramer}, where the non-covalent interactions are van der Waals as well as N$\hdashrule[0.5ex]{10mm}{.5pt}{1.0mm}$H hydrogen bonding. 
This system is small enough that will allow us to check the accuracy of the embedding  scheme by comparing it to the result of the full EOM-CC calculation.

To set up an embedding problem we must separate the orbitals present in the tetramer of ammonia into appropriate correlation domains. In order to do that, we have investigated two routes: \\ 
\begin{itemize}
    \item in the first approach, we localize all molecular orbitals of the system to approximately identify orbitals belonging to each of the monomers. Then, we consider all those orbitals belonging to a monomer as one impurity ($A_i$). There are four such impurities in the total system. It is important to mention the localization procedure we have used here. For occupied HF orbitals, existing localization procedures are well behaved, however when virtual orbitals are very diffused, most of the localization procedures are not robust enough to yield good localized results. Here, we have used a procedure proposed by H{\o}yvik and J{\o}rgensen \textit {et. al.} \cite{JorgensenJCP2013}, where the power of fourth moment of orbitals spread $\mu_4^p = \langle p | (\textbf{r} - \langle p | \textbf{r} | p \rangle)^4 | p \rangle$ is minimized. This approach tends to show overall better localization of virtual orbitals in comparison to a Pipek-Mezey or Boys localization procedure \cite{JorgensenJCP2013}. We export these localized orbitals from \texttt{LSDALTON} \cite{LSDALTON} quantum chemistry package, and then visualize them in \texttt{Chimera}\cite{Chimera2004} visualization program. The selection of impurities are done entirely based on this visualization. This scheme results in 4 impurities, each of these impurities has 13 localized orbitals and 22 bath orbitals.
\item the other approach is based on the natural orbital occupancy from the weakly correlated method of our choice and does not assume any locality of the orbitals. Here we choose a cut-off: 0.01 $<$ Occ $<$ 1.98, and consider all the orbitals within that range as one impurity. With that choice we treat 28 natural orbitals in one impurity. This impurity requires 54 bath orbitals.
\end{itemize}

The second important concern is the choice of a weakly correlated method for treating the environment. The simplest and obvious choice is HF. However, we discovered that our results with an environment treated by HF were not accurate enough. Specifically, when we have chosen to build the impurity using localized HF orbitals, we obtained not only very poor total energy and self-energy, but also poor spectral function. 
In the remaining discussion, SEET($\rm method_{env}$/$\rm method_{sys}$) denotes a SEET calculation with $\rm method_{env}$  used for the treatment of the environment while $\rm method_{sys}$ is employed to treat the impurity containing system orbitals.
The total energy in SEET(HF/CCSD) calculation differs from the CCSD energy of the total system by 0.603 mH and occupation numbers of the  highest occupied molecular orbital (HOMO) and lowest unoccupied molecular orbital (LUMO) orbitals differ by 0.027 and 0.018, respectively.   Thus, SEET(HF/CCSD) calculation with relatively small embedded fragments can yield only qualitative results. 
These errors can be attributed to the lack of dynamical correlation in the treatment of the environment.
Consequently, we decided to use a  correlated method for the description of the environment. Here, we have chosen to use GF2~\cite{Dahlen05,Zgid14,Rusakov16,Phillips15,Kananenka15,Kananenka16,Welden16,kananenka_hybrif_gf2,Iskakov_Chebychev_2018} for this purpose, however other choices such as GW~\cite{GWSEET2017} are also possible. Moreover, in general one could use a high level CC method such as CCSDTQ for the treatment of the embedded orbitals while a low level CC (e.g. CCSD) to treat the environment.     

Here, we concisely summarize in a step-wise manner how a Green's function CCSD solver has been integrated in the framework of SEET. 
For further details of the SEET algorithm we refer readers to Lan \textit{et al.}\cite{SEET16}. HL stands for a high level loop and LL denotes a low level loop. While the convergence of the low level loop is necessary, the convergence of the high level loop is optional. SPEC denotes a step in which the spectral functions are evaluated.
\begin{description}
    \item [HL1] Evaluate a GF2 Green's function, $G_{weak}$ for the total system in the AO basis.
    \item [HL2] Transform $G_{weak}$ from the AO basis to an orthogonal basis or to a natural orbital (NO) basis yielding $G^{ortho}_{weak}$. Extract a subset Green's function  $G^{A_i}_{weak}$ for each of the impurities $A_i$ from  the entire $G^{ortho}_{weak}$. 
    \item [LL1] Calculate hybridization function $\Delta^{A_i}$ for each of the impurities according to Eq. \ref{Eq:gimp}.
    \item [LL2] Build the impurity Hamiltonian, after fitting $\Delta^{A_i}$ with system-bath coupling integrals according to Eq. \ref{Eq:hybridization}.
    \item [LL3] Calculate the double-counting correction terms $[\Sigma(\omega)]_{DC}^{A_i}$ and $[\Sigma_\infty]_{DC}^{A_i}$ for each of the impurities using GF2. 
    \item [LL4] Determine the number of particles in each impurity by identifying the minimum energy solution at the CCSD level. Since for GFCC we do not have a direct diagrammatic expansion of the self-energy, for each of the impurities, we first evaluate the CCSD Green's function $G^{A_i}_{imp}$ and then evaluate $\Sigma_{imp}^{A_i}$ via Dyson equation with $G^{A_i}_0$. Here, $G^{A_i}_0$ is defined as $G^{A_i}_0=[(i\omega+\mu)\mathbf{1}-t^{A_i}-\Delta^{A_i}]^{-1}$, where for a given impurity $A_i$ $t^{A_i}=F^{A_i}-[\Sigma_\infty]_{DC}^{A_i}$ is a subset of the Fock matrix with a double counting correction subtracted. 
    \item [LL5] For each of the subsets $A_i$ build the total self-energy as $\Sigma^{A_i}_{total} = \Sigma_{weak} + \Sigma^{A_i}_{imp} - \Sigma^{A_i}_{DC}$.
    \item [LL6] Calculate a new $G_{total}$ using $\Sigma^{A_i}_{total}$ in the Dyson equation and adjust the chemical potential $\mu$ to obtain correct number of particles. For every impurity construct the subset Green's functions $G^{A_i}$ from $G_{total}$.
    \item [LL7] Repeat LL1-LL6 and test for convergence either on $\Delta^{A_i}$ or on the total energy. 
    \item [HL3] Using the $G_{total}$ evaluated in step LL6 perform a single iteration of GF2 from HL1 and continue to steps LL1-LL7. An overall convergence is achieved when the total Green's function and the electronic energy stop to change.
    \item [SPEC] Perform the analytic continuation of Tr($G_{total}$) from the imaginary frequency axis to the real frequency axis to get spectral functions. 
\end{description}

\subsubsection{Green's function CC vs EOMCC}

\begin{figure*} [htb]
\includegraphics[width=\columnwidth]{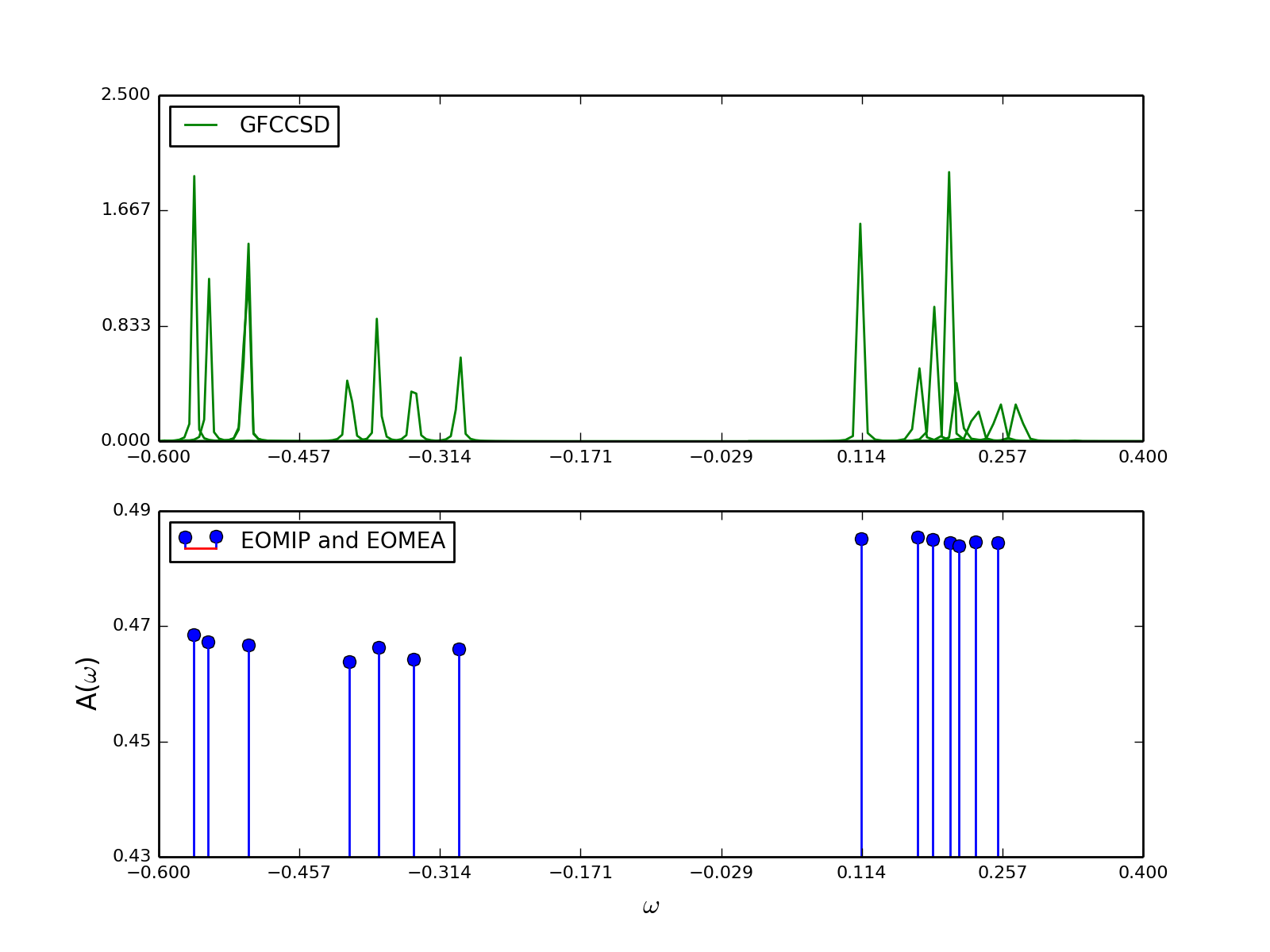} \caption{Individual orbital contributions towards the total spectral function. Total 14 orbitals have been chosen around HOMO and LUMO. We have chosen a broadening of 0.005 H to calculate the Green's function.}
\label{fig:spectra}
\end{figure*}

For (NH$_3$)$_4$, we first evaluated a spectral function of the total system at the CCSD level. This is done by performing an evaluation of Green's function from Eq.~\ref{Eq:contfrac} using real frequencies with a small imaginary broadening of 0.005 H. Subsequently, a spectral function is evaluated using Eq.~\ref{eq:spectralf}.
To compare it with EOMCC spectra, we have plotted contributions of 14 orbitals around HOMO and LUMO to the total spectral function in Fig. \ref{fig:spectra}. To evaluate intensity from EOMCC, we have calculated Dyson orbitals $\phi_d(p)$ as $\langle \Psi_N | p^\dagger | \Psi_{N-1}\rangle$. The intensity of a particular quasi-particle ionization can then be estimated as I $\propto$ $\sum_p |\phi_d(p)|^2$. We observe that the IP and EA poles obtained from the  CCSD Green's function exactly correspond to EOMCCSD poles. Moreover the intensity trend is also similar in both spectra. 

\subsubsection{Total energy of SEET with Green's function CCSD}


 Using Galitskii-Migdal formula from Eq.~\ref{Eq:GM-formula}, we have evaluated the total energy of (NH$_3$)$_4$ from GF2 and SEET(GF2/CCSD) with different choice of the impurity orbitals. In all the Green's function based approaches we have have chosen the frequency and time grid at a sufficiently low inverse temperature ($\beta = 100$ 1/a.u.). Since (NH$_3$)$_4$ is a finite system with a very large band gap, we expect that this inverse temperature is low enough to be comparable with zero temperature energies. 
 We use Green's function CCSD energy evaluated from Galitskii-Migdal formula with no frozen occupied and virtual orbitals as a benchmark energy.
 When we treat the full system with self-consistent GF2 method we get a deviation of 50 mH from the benchmark energy. The SEET(GF2/CCSD) scheme with localized orbitals as impurity basis reduces the difference from CCSD to 22 mH. When considering natural orbitals from one-shot GF2 calculation as impurity orbitals, we obtain much better energy; now the difference to CCSD is only 9 mH. 

\subsubsection{Occupation numbers of SEET with Green's function CCSD}

We have analyzed the one-body density matrix of the total system by comparing natural orbital occupancies between various methods. We summarize our results in Tab.~\ref{Tab:seet_density}. We observe that SEET significantly improves upon a single iteration of the GF2 method. The remaining differences can be even further reduced by performing the high level loop (HL) until convergence in SEET(GF/CCSD) with natural orbitals as the impurity basis. In that case, due to the iterative redefinition of the impurity orbitals, the difference for HOMO reduces to 0.0019 and for LUMO to 0.00002.   

\begin{table} [htb]
\caption{Occupation numbers obtained using different orbital bases for impurities in SEET. $\Delta$ stands for an absolute difference from the wave function based CCSD method. Both SEET (LMO) and SEET (NO) signify SEET (GF2/CCSD) calculations that differ in the choice of impurity basis. LMO $\equiv$ Localized Molecular Orbitals and NO $\equiv$ Natural Orbitals.} \label{Tab:seet_density}
\begin{tabular}{|c|c|c|c|}
\hline 
\#orbital & $\Delta$GF2 (1-shot) & $\Delta$SEET (LMO) & $\Delta$SEET (NO) \\
\hline 
\hline 
HOMO-1 & 0.0051 & 0.0022 & 0.0024 \\
\hline 
HOMO & 0.0048 & 0.0018 & 0.0026 \\
\hline 
LUMO & 0.0032 & 0.0022 & 0.0032 \\
\hline 
LUMO+1 & 0.0035 & 0.0025 & 0.0028 \\
\hline 
\end{tabular}
\end{table}

\subsubsection{Self-energy of SEET with Green's function CCSD}

We have also compared individual self-energy elements evaluated in SEET to the self-energy evaluated from the CCSD Green's function for the entire system. To facilitate such a comparison, we first calculated the SEET self-energy with different orbitals chosen for the impurity basis, specifically localized MO and NO. Subsequently, the  self-energies obtained were transformed to a canonical MO basis. Similar transformation has been carried out for the GF2 self-energy as well. The self-energy from CCSD was directly computed in MO basis. We plot self-energies  for HOMO and (HOMO-1) orbitals in Fig. \ref{fig:sigma-seet}. We observe that the self-energy for SEET with NO basis for the impurity presents a significant improvement over the GF2 self-energy and is close to the CCSD self-energy for the total system. The self-energy evaluated from SEET with the LMO basis is quite far from the CCSD self-energy for the total system and does not present a significant improvement over the GF2 self-energy. 
These stark differences in the performance of SEET schemes with different orbitals chosen to the impurity can be attributed to the missing correlation between impurities in the scheme with LMO. 
\begingroup
\centering
\begin{figure*}[]
\centering
\subfigure{\includegraphics[width=0.48\textwidth]{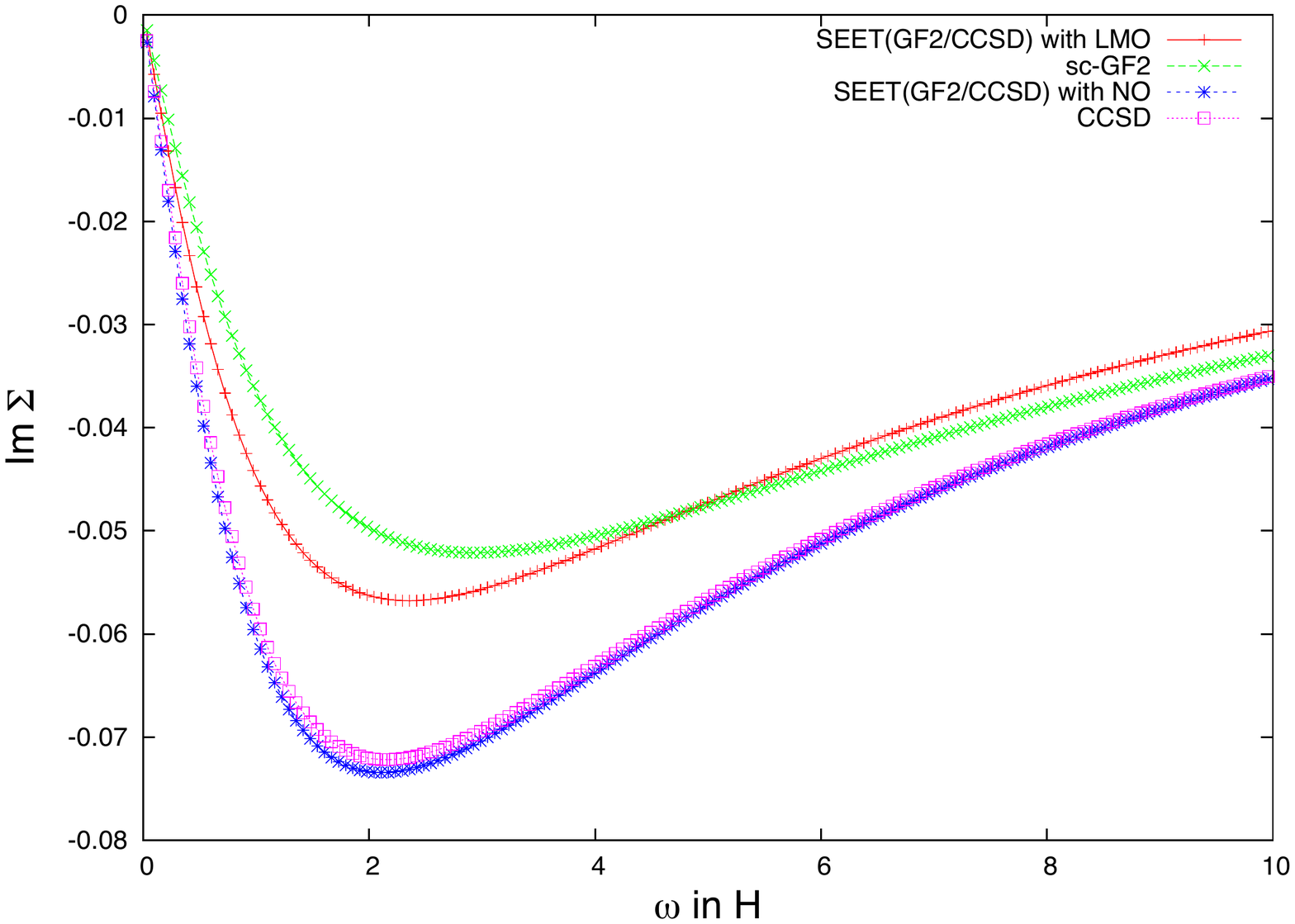}}
\subfigure{\includegraphics[width=0.48\textwidth]{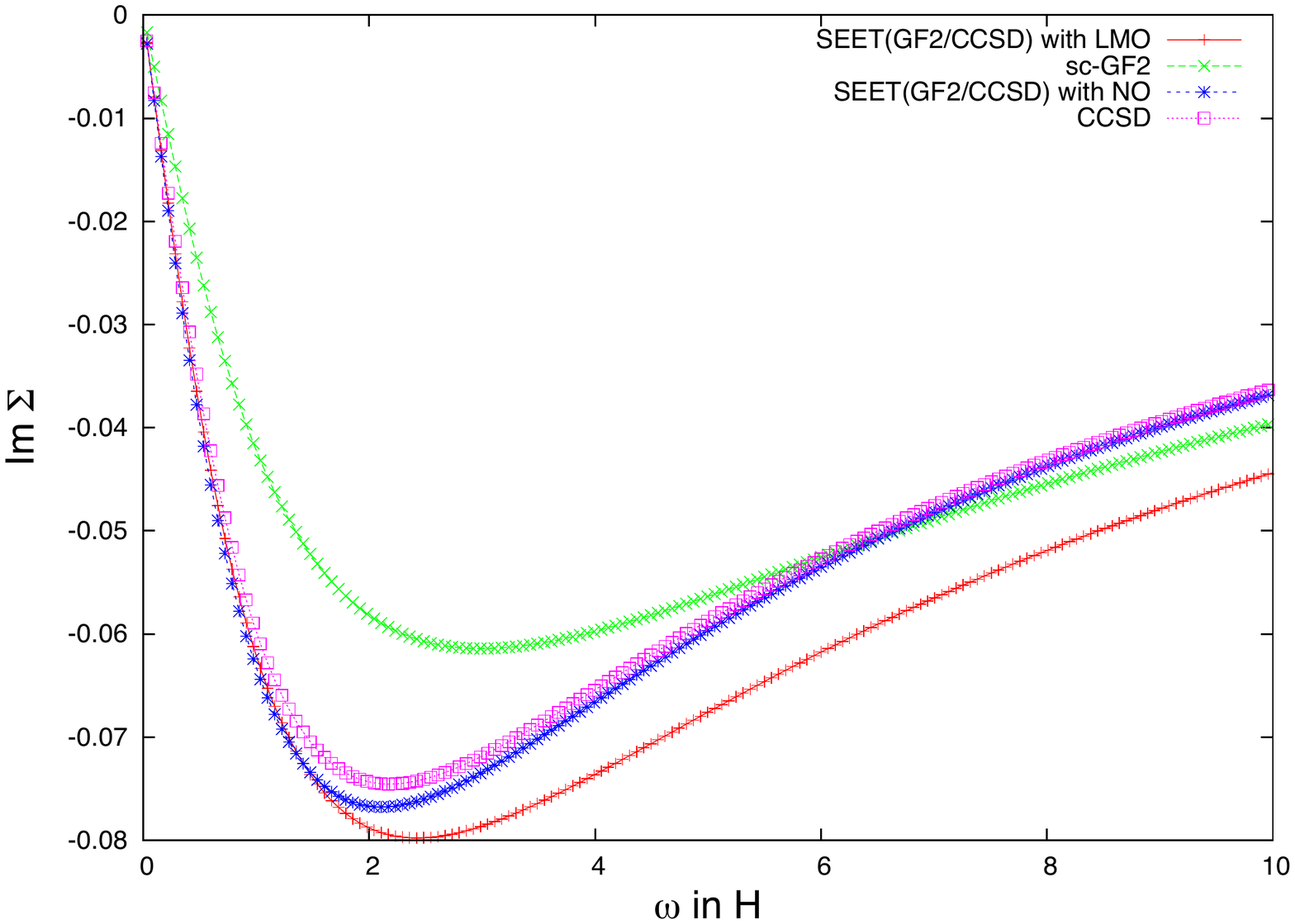}}
\caption{Comparison of self-energy from the Green's function CCSD for the whole system with SEET self-energies with different choices of impurity orbitals and also with self-consistent GF2. Left: self-energies for HOMO orbital. Right: self-energies for (HOMO-1) orbital.} \label{fig:sigma-seet}
\end{figure*}
\endgroup

\subsubsection{Spectral function of SEET with Green's function CCSD}

Finally, we compare spectral functions between Green's function CCSD for the entire problem and SEET. While the evaluation of the spectral function is usually not desired for molecular systems, this quantity is important for periodic systems.  Since the ammonia cluster is a  finite system, here, we  will only compare the HOMO-LUMO gap obtained from the full system CCSD calculation to the one obtained from SEET. 

The Green's function in SEET is calculated on the imaginary frequency axis to enable a smooth convergence of the iterative SEET procedure. However, a spectral function from Eq.~\ref{eq:spectralf}
is defined on a real frequency axis. Consequently, to obtain the real axis Green's function, an analytical continuation of imaginary frequency data is required. The analytical continuation is a challenging problem and we recommend following Ref.~\onlinecite{LEVY2017} for details. In this work, we use $\tt{Maxent}$ \cite{LEVY2017} as the continuation method and Tr($G(i\omega)/n)$, where $n$ is number of orbitals, as our input data. 

The CCSD Green's function for the entire system can be evaluated directly on the real frequency axis, thus avoiding the need for the analytical continuation procedure. In the left panel of Fig.~\ref{fig:sp_seet}, we compare the CCSD spectral function with the ones coming from SEET calculations. In the hole part of the spectral function, it is apparent that we miss many spectral features and the main peak has a broad shoulder for SEET as opposed to CCSD. The particle part of the spectrum with SEET also does not exhibit all the spectral features of CCSD but the main peak position appears to be quite accurate. 

To investigate if these discrepancies are arising from the analytical continuation errors or true errors due to the SEET procedure, we have calculated the CCSD Green's function on the imaginary frequency axis and continued it in the same manner as SEET Green's functions. In the right panel of Fig. \ref{fig:sp_seet}, we show the resulting spectral functions. The analytically continued CCSD spectral function misses some of the details of the original CCSD yielding a spectrum that compares very well with SEET. Thus, we conclude that most of the discrepancies between SEET and CCSD were due to the analytical continuation procedure.
\begingroup
\centering
\begin{figure*} [htb]
\subfigure{\includegraphics[width=0.48\textwidth]{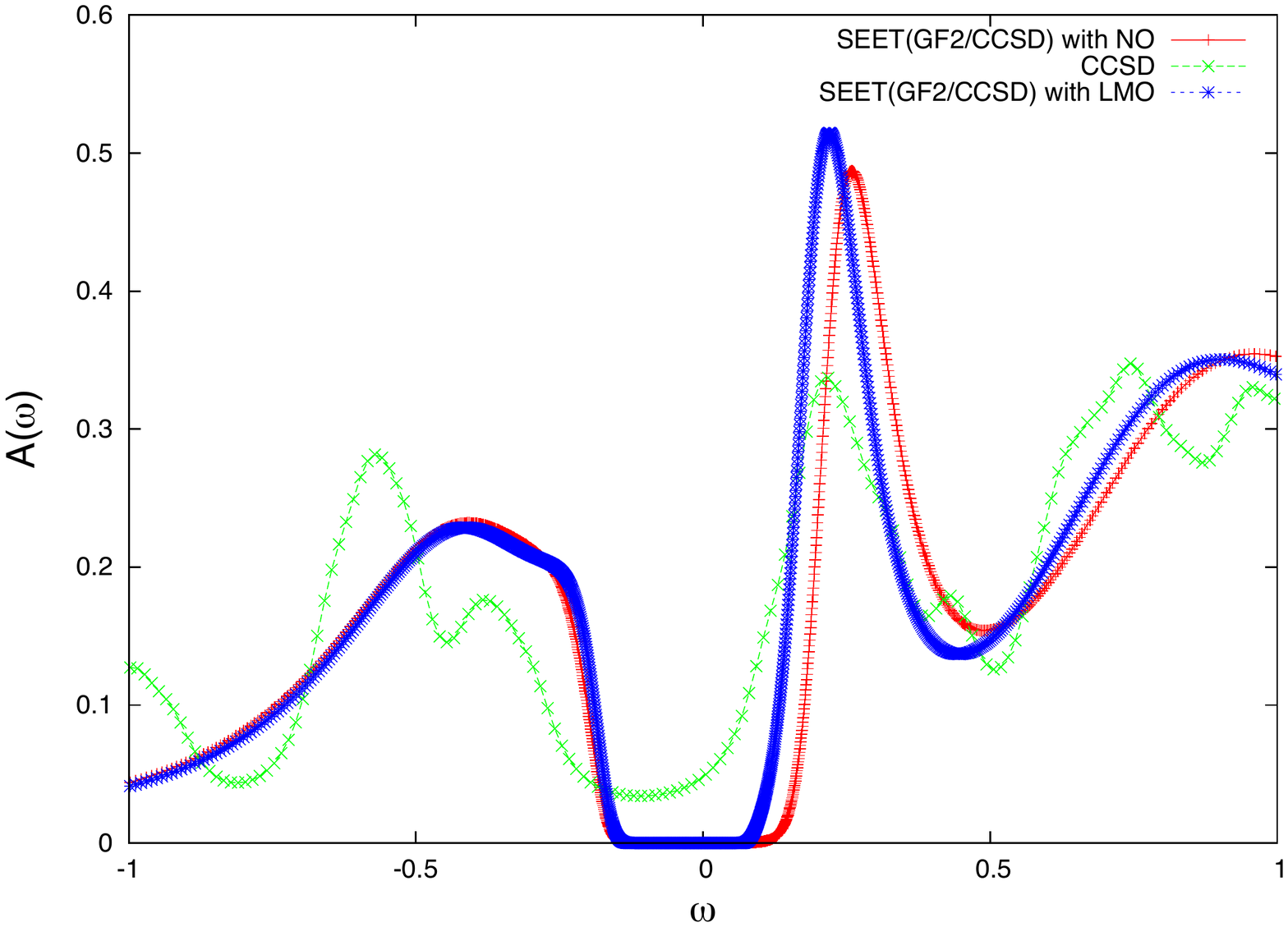}}
\subfigure{\includegraphics[width=0.48\textwidth]{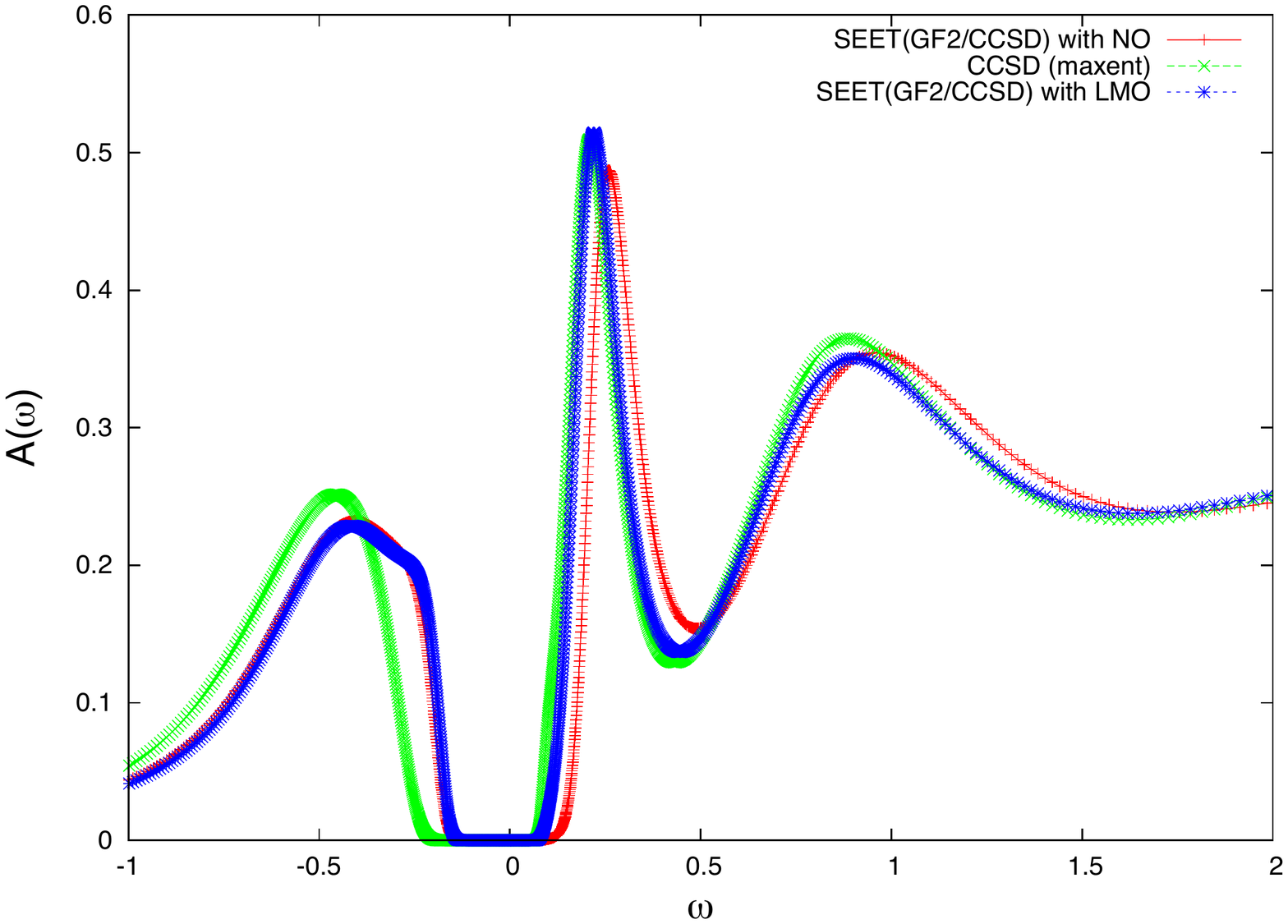}}
\caption{Comparison of spectral functions from SEET, GF2, and CCSD for the entire system. Unit of $\omega$ is in Hartrees. Left: directly calculated CCSD spectral function on the real frequency axis is compared to the analytical continuation data for SEET. Right: both CCSD and SEET spectral function were obtained by analytically continuing imaginary frequency data.}
\label{fig:sp_seet}
\end{figure*}
\endgroup

\section{Conclusion}\label{sec:conclusion}
We have outlined how CC Green's functions can be used as a solvers for Green's function embedding methods. To evaluate the CC Green's function, we used the Lanczos-based inversion of a large non-hermitian matrix. 
This algorithm has a very advantageous computational cost since it scales only as the parent CC calculation and does not depend on the number of frequency points (N$_\omega$) used in the calculation. Moreover, once the Lanczos tridiagonal matrix is known, both the Green's function on the real or imaginary axis can be calculated only be changing the frequency argument in the continued fraction expression (Eq.~\ref{Eq:contfrac}). 
The use of the Lanczos solver is also advantageous in terms of memory since only the right and left Lanczos vectors from the current and previous iterations need to be stored. These aspects of the Lanczos based CC Green's function algorithm make it optimal for embedding problems with impurities containing many orbitals where large frequency grids can be necessary.

To calibrate the CCSD Green's function solver, we have applied it to impurity problems constructed for the 1D and 2D Hubbard with different interaction strengths $U/t$.
For the 1D Hubbard model, we observed that the CCSD Green's function solver yields spectral functions and self-energies that excellently agree  with FCI for all interaction strengths analyzed (up to $U/t=8$). Moreover, the quality of those quantities from CCSD is much better than form truncated CI (CISD, CISDT) methods. The reason for this excellent agreement between FCI and CCSD is that the impurity and bath orbitals for the 1D Hubbard model have up to two unpaired electrons, thus, being an ideal case for CCSD. 
Testing for impurities (2$\times$2 cluster and 8 bath orbitals) built for the 2D Hubbard model required unrestricted CC formalism to account for all possible open-shell configurations. While in the weakly correlated regime CCSD performance was excellent, in the strongly correlated regime, we observed that the CCSD self-energy displayed significant deviations from FCI. This can be explained by analyzing occupations of the impurity and bath orbitals which show more than two unpaired electrons.
Consequently, we conclude that while the impurity problem alone seem to be less correlated than the entire problem, CC solvers with increasing excitation levels are most likely necessary to describe strongly correlated problems with increasing number of orbitals in the Anderson impurity model. This finding is in agreement to quantum chemical observations where high excitations  are necessary to describe multiple unpaired electrons that appear as a result of strong correlations. 
Consequently, we recommend the CCSD Green's function solver to treat weakly to moderately correlated problems. 


To test the applicability of GFCC as a solver for realistic embedding problems, we benchmarked it on an ammonia cluster, (NH$_3$)$_4$, that can sever as a prototype for a molecular crystal. For (NH$_3$)$_4$, we observed that neglecting correlation present in the environment and treating it only at the HF level can lead to significant errors. When GF2 is used as a low cost, perturbative method for the environment,  a significant amount of correlation present in the system is recovered leading to a very good SEET(GF2/CCSD) total energy, occupation numbers, self-energy and spectral function when compared with CC calculation of the entire problem. 

Moreover, the choice of impurity basis makes a significant difference in the accuracy of final quantities. We have first tested SEET with impurity orbitals constructed from localized molecular orbitals from HF.
This choice appeared not to be very accurate for almost  all the quantities, except occupation numbers. This indicates that, when impurities are expressed in the localized orbital basis, a significant portion of the inter-monomer correlation is missing. 
Finally,  we have chosen GF2 natural orbitals as another choice for the impurity basis. This significantly improves upon LMO basis. The total energy is only 9 mH less than the CCSD energy for the full system. Other calculated quantities are also in excellent agreement with the CCSD calculation. 

The results of the impurity and SEET benchmarks conducted here indicate that using CCSD as a solver is advantageous since it enables treatment of large impurities that are necessary to describe complex chemical problems displaying weak to moderate correlations and are too large to be treated by a brute force CC calculation on the entire problem. This observation opens an interesting research direction where traditional quantum chemistry variety of CC codes with little or no modification can be immediately used for treating solids or molecules that were previously inaccessible. More studies are necessary to investigate how GFCC behaves when both systems and environment are treated at the CC level with different excitation levels (e.g SEET(CCSD/CCSDT(Q)), how to reduce the size of the impurity by using different orbital bases, or how to accurately deal with fitting a large number of bath orbitals, and how to use generalized SEET approaches to treat overlapping impurities.

\section{Acknowledgements}
This work was supported by the Center for Scalable, Predictive methods for Excitation and Correlated phenomena (SPEC), which is funded by the U.S. Department of Energy (DOE), Office of Science, Office of Basic Energy Sciences, the Division of Chemical Sciences, Geosciences, and Biosciences. 
D.Z. thank Karol Kowalski for helpful discussions about the Lanczos-based EOM-CC scheme.
We thank Devin Matthews (SMU, Texas) for his Aquarius code that was used as a platform for our Green's function CC code. We are also grateful  to Alexander Rusakov for his help with the analytic continuation procedure. We also would like to thank Lin Lin and Leonardo Zepeda for discussions concerning fitting hybridizations with a large number of bath orbitals.
\appendix
\section{Transformation of Trial Vectors} \label{Sec:appx1}

Right hand transformed vectors: $b = \overline{a}_p$ \\
Left hand transformed vectors: $e = (1+ \Lambda) \overline{a_p^\dagger}$

Note that, we have not mentioned the rank of the b and e operators. We have chosen up to double excitation/de-excitation level of T and $\Lambda$, which from diagrammatic expressions generate up to double excitation and de-excitation level of b and e, respectively.

\subsection{$G^{h}_{pq}$}
Let us first consider that $p = q$. When, p $\in$ V
\begin{align}
b_i = {} & t^e_i a_e \nonumber \\
b^a_{ij} = {} & t^{ae}_{ij} a_e \nonumber \\
e^i = {} & l^i_e (a^\dagger)^ e \nonumber \\
e^{ij}_{a} = {} & l^{ij}_{ae} (a^\dagger) ^e \label{Eq:IP_V}
\end{align}

and, when p$\in$ O

\begin{align}
b_i = {} & a_i \nonumber \\
e^i = {} & (a^\dagger)^i - D^i_m (a^\dagger)^m \nonumber \\
e^{ij}_{a} = {} & l^i_a (a^\dagger)^j - D^{ij}_{am} (a^\dagger)^m \label{Eq:IP_O}.
\end{align}
Now, for the $p \neq q$ cases, as we have mentioned before we do calculate G$_{p+q,p+q}$ elements. Here, we have to distinguish among 4 different possibilities : OO, OV, VO and VV. Depending on the index type - occupied or virtual we carry out the same transformations as in equations A1 and A2. Afterwards, we add the corresponding contributions to the b and e vectors. For example, let us consider the OV situation:

\begin{align}
b_i = {} & a_i + t^e_i a_e \nonumber \\
b^a_{ij} = {} & t^{ae}_{ij} a_e \nonumber \\
e^i = {} & (a^\dagger)^i - D^i_m (a^\dagger)^m + l^i_e (a^\dagger)^ e \nonumber  \\
e^{ij}_{a} = {} & l^i_a (a^\dagger)^j - D^{ij}_{am} (a^\dagger)^m + l^{ij}_{ae} (a^\dagger)^e
\end{align}
where,
\begin{align}
    D^i_j = {} & l^i_e t^e_j + \frac{1}{2} l^{im}_{ef} t^{ef}_{jm} \nonumber \\
    D^{ij}_{am} = {} & l^{ij}_{ae} t^e_m 
\end{align}
Rest of the types will follow the same procedure.
\subsection{$G^e_{pq}$}
 For $p=q$ cases, when $p \in V$ 
\begin{align}
b^a = {} & a^a \nonumber \\
e_a = {} & (a^\dagger)_a + (a^\dagger)_e D^e_a \nonumber \\
e^{i}_{ab} = {} & l^i_a (a^\dagger)_b + D^{ei}_{ab} (a^\dagger)_e
\end{align}
and when $p \in O$
\begin{align}
b^a = {} & t^a_k a^k \nonumber \\
b^{ab}_i = {} & - t^{ab}_{ik} a^k \nonumber \\
e_a = {} & -l^k_a (a^\dagger)_k \nonumber \\
e^i_{ab} = {} & -l^{ik}_{ab} (a^\dagger)_k.
\end{align}

where,
\begin{align}
    D^a_b ={} & -l^m_b t^a_m - \frac{1}{2} l^{km}_{be} t^{ae}_{km} \nonumber \\
    D^{am}_{ef} = {}& -l^{nm}_{ef}t^a_n 
\end{align}

For $p \neq q$, we will follow the same procedure as G$^h$.

\end{document}